\documentclass[aps,preprint,superscriptaddress,showpacs]{revtex4}
\usepackage{graphicx}
\usepackage{dcolumn}
\usepackage{amsmath}

\begin{document}
\bibliographystyle{apsrev}

\title{Acceleration effect of coupled oscillator systems}
 
\author{Toru Aonishi}

\affiliation{Brain Science Institute, RIKEN, 
2-1 Hirosawa, Wako-shi, Saitama, 351-0198, Japan 
}

\author{Koji Kurata}

\affiliation{Department of Mechanical Systems Engineering, Faculty of Engineering, 
University of the Ryukyus, Sembaru 1, Nishihara, Okinawa 903-0213, Japan
}

\author{Masato Okada}

\affiliation{ERATO Kawato Dynamic Brain Project, 2-2 Hikaridai, Seika-cho, 
Soraku-gun, Kyoto 619-0288, Japan
}
\affiliation{Brain Science Institute, RIKEN, 
2-1 Hirosawa, Wako-shi, Saitama, 351-0198, Japan 
}
\date{\today}

\begin{abstract} 

We have developed a curved isochron clock (CIC) by modifying the radial
isochron clock to provide a clean example of the acceleration
(deceleration) effect. By analyzing a two-body system of coupled CICs,
we determined that an unbalanced mutual interaction caused by curved
isochron sets is the minimum mechanism needed for
generating the acceleration (deceleration) effect
in coupled oscillator systems. From this we can see that the 
Sakaguchi and Kuramoto (SK) model which is a class of 
non-frustrated mean feild model has an acceleration 
(deceleration) effect mechanism. To study frustrated coupled
oscillator systems, we extended the SK model to two oscillator
associative memory models, one with symmetric and one with asymmetric
dilution of coupling, which also have the minimum mechanism 
of the acceleration (deceleration) effect. We theoretically found that
the {\it Onsager reaction term} (ORT), which is unique to frustrated
systems, plays an important role in the acceleration (deceleration)
effect. These two models are ideal for evaluating the effect
of the ORT because, with the exception of the ORT, they have the same
order parameter equations. We found that the two models have
identical macroscopic properties, except for the acceleration effect
caused by the ORT. By comparing the results of the two
models, we can extract the effect of the ORT from only the
rotation speeds of the oscillators.
\end{abstract}
\maketitle

\section{Introduction}

Coupled oscillators are of intrinsic interest in many branches of
physics, chemistry, and biology. One class of coupled oscillator systems
has a property that by mutual interactions, the oscillatory frequency of
an individual unit is made higher (lower) than its natural
frequency. This phenomenon is called the "acceleration (deceleration)
effect" and is of particular interest to researchers in the biological
branch of mathematics \cite{meunier,Hansel1}. However, we still do not
have a clear understanding of the acceleration (deceleration) effect; we
need to clarify the basic mechanism of this effect in coupled oscillator
systems.

In the first part of this paper, we treat general oscillator models
coupled weakly by general coupling terms according to Ermentrout
\cite{ermentrout}, and we derive one-dimensional phase equations from
original equations of high-dimensional dynamics. Then, we apply this
general method to the radial isochron clock (RIC), which has very simple
oscillator dynamics on $\bf{R}^2$, i.e., a unit circle stable orbit.
Next, we develop a curved isochron clock (CIC) by modifying the RIC, 
and we derive one-dimensional phase equations from coupled CICs. 
The CIC also has a unit circle stable orbit. We demonstrate that the CIC is a
very simple model that provides a clean example of the acceleration
(deceleration) effect caused by diffusion coupling. Our analysis shows
that the Sakaguchi and Kuramoto (SK) model \cite{kuramoto}, which is a
mean field model of coupled oscillators, has the minimum mechanism of
the acceleration (deceleration) effect deeply related to coupled
CICs. The SK model is not frustrated, so we need to study how frustrated
interactions affect the frequency of oscillator systems.

In the next part of this paper, we propose a mean field theory that can
treat a general class of frustrated coupled oscillator systems and use
it to clarify the mechanism of the acceleration (deceleration) effect in
frustrated coupled oscillators. We found that the {\it Onsager reaction
term} (ORT), which describes the effective self-interaction, plays a key
role in the effect. The ORT is of great importance in obtaining a physical  
understanding of frustrated random systems, because the 
presence of such an effective self-interaction is one of 
the characteristics that distinguish frustrated and non-frustrated
systems of this types. For equilibrium systems,
we can rigorously evaluate the effect of the ORT by using the
Thouless-Anderson-Palmer (TAP) framework \cite{Mezard} and/or by using
the replica method \cite{fukai2}. However, we cannot directly apply
these systematic methods to non-equilibrium coupled-oscillator
systems. While we can define a formal Hamiltonian function on such
systems, Perez and Ritort demonstrated that the ground states of such a
Hamiltonian are not stationary states of the dynamics
\cite{Perez}. Therefore, it is impossible to construct a theory based on
free energy for such systems. Consequently, to evaluate the macroscopic
quantities in such systems that include an ORT, {\it self-consistent
signal-to-noise analysis} (SCSNA), which can be applied to systems
without a Hamiltonian function, has been used \cite{fukai3}. The
mathematical treatment of this method is similar to that of the cavity
method \cite{Mezard}. Results obtained using SCSNA have been consistent
with those using the replica method, but this method 
includes a few heuristic steps. While SCSNA has produced
some interesting results, they have not been sufficient to give a
complete understanding of frustrated systems. Consequently, many
fundamental theoretical questions remain in the study of such
systems. In fact, even the existence of the type of self-interaction
that can be described by the ORT is the subject of some debate
\cite{aonishi2,yoshioka}.

Here, we consider two oscillator associative memory models, 
one with symmetric and one with asymmetric dilution of coupling. 
These two models use are ideal for evaluating the effect of the ORT
because, with the exception of the ORT, they have the same order
parameter equations. The theory we present reveals a non-trivial
phenomenon: oscillator rotation in a symmetric diluted model is faster
(slower) than that in an asymmetric diluted model, even if the two
models have identical macroscopic properties. Therefore, by comparing
the results of the two models, we can extract the effect of the ORT
from only the rotation speed of the oscillators.

As the random dilution of coupling in associative memory models
is equivalent to the random coupling
noise in the thermodynamic limit, as revealed by previously described
theories of equilibrium systems \cite{Sompolinsky,aoyagi2,okada}, 
the present model is reduced to one for glass oscillators \cite{daido} 
in the limit of strong dilution. 
Therefore, the theory we propose covers two types of frustrated 
systems, the oscillator associative memory model and the
glass oscillator model. Such models are typical frustrated
non-equilibrium systems with large degrees of freedom.

In uniformly coupled oscillators, there is a unique stable
state, i.e., the ferromagnetic state in the phase space. 
In random systems, there are many stable states in the phase
space (ferromagnetic phases and glass phases). Our theory describes the
mutual entrainment in the ferromagnetic phases (memory
retrieval), in which most of the oscillators are synchronized by
the strong mutual interaction. If the memory retrieval
process is unsuccessful, the system is in the glass
phase (spurious memory retrieval), and in this phase, the
system causes quasi-entrainment \cite{daido}, which is
regarded as weak entrainment compared to that in the ferromagnetic
phase. Unfortunately, it is difficult to theoretically analyze the glass
states of non-equilibrium systems because we have not yet developed
sufficient theoretical tools to capture the complicated structures of
the glass state in non-equilibrium systems. Therefore, instead of using
theoretical analyses, we have numerically studied quasi-entrainment in
the glass phase \cite{daido}. We found that the distribution of local
fields takes a "volcanic" form in the glass phase \cite{daido}, which
implies an outbreak of the ergodicity breaking with the ultrametric
structure of the glass state related to the replica-symmetry breaking.

A serious problem with using attractor-type networks for solving
optimization problems is detecting being trapped in a meta-stable state
during the relaxation process. Results obtained from analyzing
memory retrieval and spurious memory retrieval have shown
that it is possible to determine whether the retrieval process
is successful or not by using information about the
synchrony/asynchrony. This means that we can apply non-equilibrium
systems to optimization problems in order to detect meta-stable states.

\section{Phase equation}

In this section, we use the method of Ermentrout \cite{ermentrout} to
derive a phase equation for coupled oscillators. First, let us consider
the following isolated limit cycle oscillator:
\begin{eqnarray}
\frac{d \mbox{\boldmath$x$}}{d t} = 
\mbox{\boldmath$F$}(\mbox{\boldmath$x$}), \ \ 
\mbox{\boldmath$x$} \in {\bf R}^n,\ \ \mbox{\boldmath$F$}: {\bf R}^n 
\rightarrow {\bf R}^n.\label{eq1:md1}
\end{eqnarray}
We assume this system has a stable periodic solution 
$\mbox{\boldmath$\Phi$}(t)$ with period $2 \pi$ that satisfies
\begin{eqnarray}
\mbox{\boldmath$\Phi$}'(t) = 
\mbox{\boldmath$F$}(\mbox{\boldmath$\Phi$}(t)),\ \ \mbox{\boldmath$\Phi$}(
t) = \mbox{\boldmath$\Phi$}(t+2 \pi).\label{eq1:md2}
\end{eqnarray}
This equation is autonomous or invariant to shifts in the time domain,
so $\mbox{\boldmath$\Phi$}(t + \phi)$ is also a solution for any $\phi
\in {\bf R}/2\pi$.  In other words, the periodic solution is
irresistant to a temporal shift while it conserves a fixed
orbit (neutral stability). This is referred to as "orbit stability".
Here, $\phi$ stands for the "phase" of the periodic solution.

If we modulate the time constant,
\begin{eqnarray}
(1- \varepsilon \omega)\frac{d \mbox{\boldmath$x$}}{d t} = 
\mbox{\boldmath$F$}(\mbox{\boldmath$x$}),
\end{eqnarray}
this system has a periodic solution with period $2 \pi(1-
\varepsilon \omega)$, which can be expressed as 
\begin{eqnarray}
\mbox{\boldmath$x$}(t) = \mbox{\boldmath$\Phi$}\left(\frac{t}{1- \varepsilon 
\omega} \right),
\end{eqnarray}
where $1 >> \varepsilon > 0$.
When $\omega>0$, the period of this system is slightly shorter
than $2 \pi$.

Next, we consider the high-dimensional dynamics of coupled oscillator
systems:
\begin{eqnarray}
(1 - \varepsilon \omega_i) \frac{d \mbox{\boldmath$x$}_i}{d t} &=& 
\mbox{\boldmath$F$}(\mbox{\boldmath$x$}_i) 
+ \varepsilon \mbox{\boldmath$p$}_i, \quad i = 1, \cdots ,N, \label{eq1:org1}\\
\mbox{\boldmath$p$}_i &=& \sum_{j (\neq i)}^N
\mbox{\boldmath$V$}_{ij}(\mbox{\boldmath$x$}_i, \mbox{\boldmath$x$}_j), \\
\mbox{\boldmath$x$}_i &\in& {\bf R}^n,\ \ 
\mbox{\boldmath$V$}_{ij} : {\bf R}^n \times {\bf R}^n \rightarrow {\bf R}^n,
\end{eqnarray}
where $\mbox{\boldmath$x$}_i$ is a configuration variable of the $i$-th
oscillator (with a total of $N$ oscillators). The $\varepsilon
\mbox{\boldmath$p$}_i$ is the perturbation, i.e., the coupling term,
which is the sum of $\mbox{\boldmath$V$}_{ij}$ $(i, j=1,\cdots,N)$
representing the interaction from unit $j$ to unit $i$.  If
$\varepsilon \mbox{\boldmath$p$}_i =0$, each oscillator continues
rotating on a limit-cycle orbit individually. The $\varepsilon \omega_i$
denotes the fluctuation in the individual natural frequency.

If $\varepsilon$ is sufficiently small, the components of the
perturbation that breaks the shape of the orbit are suppressed by the
stability of the solution. However, the component of the perturbation
that shifts the phase cannot be suppressed, causing the phase to move to
the most "comfortable" position.

The solution of a perturbed system (\ref{eq1:org1}) can be represented as 
\begin{eqnarray}
\mbox{\boldmath$u$}_i(t) = \mbox{\boldmath$\Phi$}(t + 
\phi_i(\tau))+\varepsilon
\tilde{\mbox{\boldmath$u$}_i}(t), \ \ \tau = \varepsilon t, \label{eq1:ps1}
\end{eqnarray}
where $\phi_i$ is the phase of the $i-$th oscillator (with a total of
$N$ oscillators), $\tau$ denotes a slowly varying time, and
$\varepsilon \tilde{\mbox{\boldmath$u$}_i}$ is a fluctuation caused by
the perturbation.  In the following derivation, $\tau$ and $\phi_i$ are
considered to be approximately constant within a period.

By substituting Eq. (\ref{eq1:ps1}) into Eq. (\ref{eq1:org1}), expanding
a polynomial around $\varepsilon=0$, and neglecting the higher order
terms, we obtain
\begin{widetext}
\begin{eqnarray}
\mbox{\boldmath$\Phi$}'(t + \phi_i(\tau))\left(\omega_i - \frac{d \phi_i}{d 
\tau}\right) &+& \sum_{j (\neq i)}^N 
\mbox{\boldmath$V$}_{ij}\left(\mbox{\boldmath$\Phi$}(t + \phi_i(\tau)) , 
\mbox{\boldmath$\Phi$}(t +\phi_j(\tau))\right) =
L_{\phi_i} \tilde{\mbox{\boldmath$u$}_i}, \label{eq1:exp1} \\
L_{\phi_i} &=& \frac{d}{dt} - 
D\mbox{\boldmath$F$}(\mbox{\boldmath$\Phi$}(t + \phi_i)),
\label{eq1:lio1}
\end{eqnarray}
\end{widetext}
where $L_{\phi}$ is the linearized operator of Eq. (\ref{eq1:md1})
around the periodic solution $\mbox{\boldmath$\Phi$}(t + \phi)$, and
$D\mbox{\boldmath$F$}(\mbox{\boldmath$\Phi$})$ is a Jacobi matrix of
$\mbox{\boldmath$F$}(\mbox{\boldmath$\Phi$})$.  A linearized slow
dynamics around the periodic solution, $\mbox{\boldmath$\Phi$}(t +
\phi)$, is expressed as 
\begin{eqnarray}
\frac{\partial \mbox{\boldmath$u$}}{\partial \tau} = L_{\phi} 
\mbox{\boldmath$u$},
\end{eqnarray}
where all of the eigen-values of $L_{\phi}$ are non-positive since the
solution, $\mbox{\boldmath$\Phi$}(t + \phi)$, is stable.  We obtain
eigenvalue $0$ of $L_{\phi}$ with eigenfunction
$\mbox{\boldmath$\Phi$}'(t + \phi)$ by differentiating $\frac{d
\mbox{\boldmath$\Phi$}}{dt} =
\mbox{\boldmath$F$}(\mbox{\boldmath$\Phi$})$.  This eigenfunction
corresponds to the minimal temporal shift because
$\mbox{\boldmath$\Phi$}(t + \phi+\varepsilon) \stackrel{\cdot}{=}
\mbox{\boldmath$\Phi$}(t + \phi) + \varepsilon \mbox{\boldmath$\Phi$}'(t
+ \phi)$.  We assume there are no other eigenfunctions for eigen-value
$0$ in the space of the periodic function, so,
\begin{eqnarray}
{\rm ker} L_\phi = {\rm span} \{ \mbox{\boldmath$\Phi$}'(t + \phi) \}. 
\label{ee8}
\end{eqnarray}
This assumption is equivalent to that for the orbit stability of 
$\mbox{\boldmath$\Phi$}(t + \phi_i)$.

We define an inner product of two $n$-dimensional $2\pi$-periodic functions as
\begin{eqnarray}
\left<\mbox{\boldmath$v$}_1(t), \mbox{\boldmath$v$}_2(t)\right> = \int_{0}^{2 
\pi} dt \mbox{\boldmath$v$}_1(t)^T \mbox{\boldmath$v$}_2(t). \label{ee9}
\end{eqnarray}
The adjoint operator, $L^{\ast}_{\phi}$, of $L_{\phi}$ is defined by 
\begin{eqnarray}
\left< \mbox{\boldmath$u$}_1,L_{\phi}\mbox{\boldmath$u$}_2 \right>
= \left< L^{\ast}_{\phi} \mbox{\boldmath$u$}_1,\mbox{\boldmath$u$}_2 \right>. 
\nonumber
\end{eqnarray}
We can explicitly obtain the adjoint operator of $L_{\phi_i}$ as 
\begin{eqnarray}
L^{\ast}_\phi = -\frac{d}{dt} - 
D\mbox{\boldmath$F$}(\mbox{\boldmath$\Phi$}(t + \phi))^T. \label{ee10}
\end{eqnarray}
From Fredholm's alternative \cite{ermentrout}, there is 
a $\mbox{\boldmath$\Phi$}^{\ast}$ that
spans a kernel of $L^{\ast}_\phi$ in the space of the periodic function, so
\begin{eqnarray}
{\rm ker} L^{\ast}_\phi = {\rm span} \{ \mbox{\boldmath$\Phi$}^{\ast}(t + \phi) 
\}. 
\end{eqnarray}

Taking the inner product between $\mbox{\boldmath$\Phi$}^{\ast}(t+
\phi_i)$ and Eq. (\ref{eq1:exp1}), we obtain
\begin{widetext}
\begin{eqnarray}
\left<\mbox{\boldmath$\Phi$}^{\ast}(t+ \phi_i), \mbox{\boldmath$\Phi$}'(t + 
\phi_i)\right>
\left(\omega_i-\frac{d \phi_i}{d \tau}\right) &+& \sum_{j (\neq i)}^N
 \left<\mbox{\boldmath$\Phi$}^{\ast}(t+ \phi_i), 
\mbox{\boldmath$V$}_{ij}\left( \mbox{\boldmath$\Phi$}(t + 
\phi_i),\mbox{\boldmath$\Phi$}(t + \phi_j)\right)\right> \nonumber \\
&=& \left<\mbox{\boldmath$\Phi$}^{\ast}(t+ \phi_i), L_{\phi_i} 
\tilde{\mbox{\boldmath$u$}_i}\right> \nonumber \\
&=& \left<L^{\ast}_{\phi_i}\mbox{\boldmath$\Phi$}^{\ast}(t+ \phi_i), 
\tilde{\mbox{\boldmath$u$}_i}\right> \nonumber \\
&=& 0. 
\end{eqnarray}
\end{widetext}
Thus, we derive the following phase equation describing the slow
dynamics of the phase-locking.
\begin{eqnarray}
\frac{d \phi_i}{d\tau} &=& \omega_i + 
\sum_{j (\neq i)}^N \Gamma_{ij}(\phi_j - \phi_i), \label{eq.org}
\end{eqnarray}
where $\Gamma_{ij}(\phi)=  \left<\mbox{\boldmath$\Phi$}^{\ast}(t),
\mbox{\boldmath$V$}_{ij}\left(
\mbox{\boldmath$\Phi$}(t),\mbox{\boldmath$\Phi$}(t +
\phi)\right)\right>/\left<\mbox{\boldmath$\Phi$}^{\ast}(t),
\mbox{\boldmath$\Phi$}'(t)\right>$. $\Gamma_{ij}(\phi)$ is referred to
as "coupling function", and $\omega_i$ represents the natural
frequency of unit $i$.  By using the formal multiple-scale perturbation
method, we reduce the high-dimensional dynamics of oscillators to a
low-dimensional representation.

\section{Acceleration effect in diffusionally coupled oscillators
(two-body system)}

In this section, we treat general oscillator models coupled weakly by
diffusional coupling terms. The general theory is applied to 
the radial isochron clock (RIC) and curved
isochron clock (CIC). Note that RIC and CIC belong to a class of 
the Stuart-Landau oscillator \cite{kuramoto0}. 
By analyzing two-body systems of coupled
RICs and coupled CICs, we clarify the general mechanism of the
acceleration (deceleration) effect in coupled oscillator systems.

We consider a system of two oscillators coupled by weak diffusion:
\begin{eqnarray}
\left\{ \begin{array}{l}
(1- \varepsilon \omega_1)\frac{d \mbox{\boldmath$x$}_1}{d t} = 
\mbox{\boldmath$F$}(\mbox{\boldmath$x$}_1) + \varepsilon 
\sigma(\mbox{\boldmath$x$}_2-\mbox{\boldmath$x$}_1), \\
(1- \varepsilon \omega_2)\frac{d \mbox{\boldmath$x$}_2}{d t} =
\mbox{\boldmath$F$}(\mbox{\boldmath$x$}_2) + \varepsilon
\sigma(\mbox{\boldmath$x$}_1-\mbox{\boldmath$x$}_2),  
\end{array} \right.\label{eq1:copsci1}
\end{eqnarray}
where $\sigma$ is the diffusion coefficient representing the coupling
strength.

Based on the analysis in Section II, we can derive the following phase
equation describing the slow dynamics of phase-locking.
\begin{eqnarray}
\left\{ \begin{array}{l}
\frac{d \phi_1}{d \tau} = \omega_1+ \sigma \Gamma(\phi_2-\phi_1), \\
\frac{d \phi_2}{d \tau} = \omega_2+ \sigma
\Gamma(\phi_1-\phi_2), 
\end{array} \right. \label{eq1:ph1}
\end{eqnarray}
where $\Gamma(\phi)= \left<\mbox{\boldmath$\Phi$}^{\ast}(t),
\mbox{\boldmath$\Phi$}(t + \phi)- 
\mbox{\boldmath$\Phi$}(t )\right>/\left<\mbox{\boldmath$\Phi$}^{\ast}(t), 
\mbox{\boldmath$\Phi$}'(t)\right>$.

In the special case, these two oscillators are
mutually locked and are accelerated (decelerated) by the effect of
diffusional coupling. We term this phenomenon "the acceleration
(deceleration) effect". We next derive the conditions for it.

We can express phase-locking solution of Eq. (\ref{eq1:ph1}) as
\begin{eqnarray}
\phi_1(\tau) = \omega \tau + \eta_1, \ \ 
\phi_2(\tau) = \omega \tau + \eta_2, \ \ 
\eta = \eta_2 - \eta_1, \label{eq1:so1}
\end{eqnarray}
where $\omega$, $\eta_1$, and $\eta_2$ are constant. When $\sigma = 0$,
the natural periods of the two oscillators are $\frac{2 \pi}{1 +
\varepsilon \omega_1}$ and $\frac{2 \pi}{1 + \varepsilon \omega_2}$,
respectively. On the other hand, when the two oscillators are mutually
locked ($\sigma \neq 0$), their periods are equal to $\frac{2 \pi}{1 +
\varepsilon \omega}$. Here, we assume 
\begin{eqnarray}
\omega_1 > \omega_2. \label{eq:as1}
\end{eqnarray}
This assumption does not lose the generality of our theory.

To obtain the parameter regions of the acceleration (deceleration)
effect, we need to define the effect. One definition for the effect is
the following. If $\omega>(\omega_1+\omega_1)/2$, the two oscillators
are locked at a speed faster than their mean natural rotation
speed. This condition is defined as "acceleration". If
$\omega<(\omega_1+\omega_1)/2$, the two oscillators are locked at a
speed slower than their mean natural rotation speed. This is defined as
"deceleration". However, in this section, we focus on more a radical
situation.

If $\omega>\omega_1$, the two oscillators are locked at a speed faster
than either of their natural rotation speeds. This condition is defined
as "acceleration". If $\omega<\omega_2$, the two oscillators are locked
at a speed slower than either of their natural rotation speeds. This is
defined as "deceleration". If $\omega_2 \leq \omega \leq \omega_1$, the
two oscillators are locked at a speed midway between their 
natural rotation speeds. This condition is called the "medium state",
and it does not belong to the acceleration (deceleration)
effect. In the following analyses of two-body systems, we use these more
radical definitions.

By substituting Eq. (\ref{eq1:so1}) into Eq. (\ref{eq1:ph1}), we obtain
\begin{eqnarray}
\omega = \omega_1 + \sigma \Gamma(\eta) = \omega_2 + \sigma \Gamma(-\eta).
\label{eq1:cond1}
\end{eqnarray}
We can rewrite Eq. (\ref{eq1:cond1}) as
\begin{eqnarray}
\Gamma(\eta) - \Gamma(-\eta) = - \frac{\omega_1 - \omega_2}{\sigma}.
\label{eq1:cond2}
\end{eqnarray}
Consequently, we can graphically obtain $\eta$ from
Eq. (\ref{eq1:cond2}) and then obtain $\omega$ from
Eq. (\ref{eq1:cond1}). Since, in general, Eq. (\ref{eq1:cond2})
possesses two or more solutions consisting of stable and unstable fixed
points, the following stability condition must be satisfied:
\begin{eqnarray}
\sigma \left(\Gamma'(\eta) + \Gamma'(-\eta)\right) > 0.\label{eq1:cond3}
\end{eqnarray}
Given Eq. (\ref{eq:as1}), we obtain the following conditions for the
acceleration (deceleration) effect from Eq. (\ref{eq1:cond1}).
\begin{eqnarray}
& & \sigma \Gamma(\eta) > 0 : \ \ {\rm (acceleration)},\label{eq1:cond4}\\
& & \sigma \Gamma(-\eta) < 0 : \ \ {\rm (deceleration)}, \nonumber
\end{eqnarray}
These conditions imply that mutual couplings between two
oscillators are asymmetric; that is, $\Gamma(\eta) \neq
\Gamma(-\eta)$. Consequently, asymmetric mutual interaction is the
essence of the acceleration (deceleration) effect.

Next, we apply this general theory to two special models: the radial
isochron clock (RIC) and the curved isochron clock (CIC). In general,
limit cycle oscillators have the so-called "isochron", which is defined
as a set of initial states converging to a oscillatory solution with a
common phase.

\begin{figure}
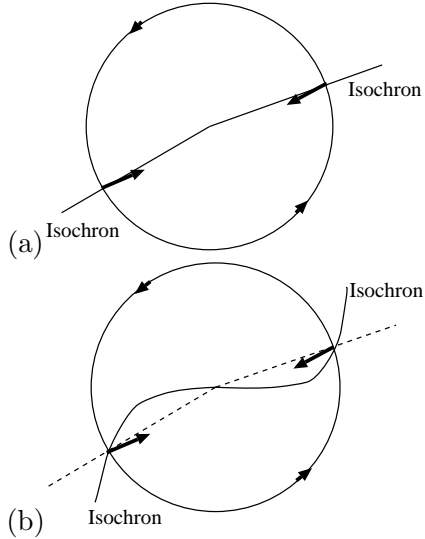

(a)\includegraphics[width=5cm]{ao_fig1a.eps}\\
(b)\includegraphics[width=5cm]{ao_fig1b.eps}
\caption{(a) Schematic diagram of two diffusionally coupled RICs and (b)
 of two 
diffusionally coupled CICs.}
\label{f1}
\end{figure}

The RIC is one of the simplest oscillators on $\bf{R}^2$, which has a
unit circle orbit with period $2 \pi$ and isochron sets that are
half-lines radiating from the origin (see Figure \ref{f1}(a)). RIC is
expressed in the polar coordinate system as 
\begin{eqnarray}
\left\{ \begin{array}{l}
\dot{r} = r (1 - r^2) \\
\dot{\theta} = 1
\end{array} \right. . \label{eq1:ric}
\end{eqnarray} 
We schematically study two-body systems with diffusional coupling; these
systems consist of a faster and a slower RIC. As shown in Figure
\ref{f1}(a), the two oscillators pull each other due to the effect of their
diffusional coupling. One is pulled backward from the isochron, while
the other is pulled forward.  As a result, one is decelerated and the
other is accelerated throughout a period. Thus, the two oscillators are
locked at a speed midway between their natural rotation
speeds.

Next, we describe our proposed curved isochron clock (CIC), which is
defined as 
\begin{eqnarray}
\left\{ \begin{array}{l}
\dot{r} = r (1 - r^2) \\
\dot{\theta} = 1 + \omega(r)
\end{array} \right. , \label{eq1:cic}
\end{eqnarray}
where $\omega(1) = 0$. The CIC has a unit circle orbit with period $2
\pi$ and curved isochron sets (see Figure \ref{f1}(b)). Figure
\ref{f1}(b) shows that if there is a phase difference, the oscillators
are pulled forward from the isochron. This happens because the isochrons
of the CIC intersect non-orthogonally with a limit cycle.  Accordingly,
the two oscillators can be accelerated throughout a period by locking
them with a phase difference. Thus, we should be able to lock two
oscillators at a faster speed than either of their natural rotation
speeds.

This consideration can be applied to the general case of weakly
diffusionally coupled oscillators. However, in general, the limit cycle
and isochron sets are not rotation symmetric, so the acceleration
(deceleration) effect must be averaged through a period.

We next derive the coupling function of diffusionally coupled CICs. A
solution on the unit circle orbit is expressed in the orthogonal
coordinate system as follows:
\begin{eqnarray}
\mbox{\boldmath$\Phi$}(t) = \left( \begin{array}{l}
\cos t \\
\sin t
\end{array} \right).
\end{eqnarray}
In this case, we can explicitly derive $\mbox{\boldmath$\Phi$}^{\ast}(t)$:
\begin{eqnarray}
\mbox{\boldmath$\Phi$}^{\ast}(t) = \frac{1}{2\pi}\left( \begin{array}{c}
-\sin t \\
\cos t
\end{array} \right) + \frac{\omega'(1)}{4\pi}
\left( \begin{array}{l}
\cos t \\
\sin t
\end{array} \right),
\end{eqnarray}
As a result, we obtain the coupling function:
\begin{eqnarray}
\Gamma(\eta) &=& \frac{1}{\cos 
\beta_0}\left(\sin(\eta+\beta_0)-\sin\beta_0\right), \label{eq1:cf1}\\
\beta_0 &=& \tan^{-1} \frac{\omega'(1)}{2}, \ \ 
-\frac{\pi}{2} < \beta_0 < \frac{\pi}{2}. \nonumber
\end{eqnarray}
Here, $\beta_0$ is derived from the intersection angle between the
isochron and the orbit. If $\beta_0=0$, Eq. (\ref{eq1:cf1}) corresponds
to weakly coupled RICs. Consequently, this phase reduction maintains the
essence of the acceleration (deceleration) effect. We can
study the acceleration (deceleration) phenomenon of diffusionally
coupled CICs by analyzing Eq. (\ref{eq1:cf1}). The parameter regions of
the acceleration (deceleration) effect, which are obtained from
the conditions defined by Eqs. (\ref{eq1:cond2}), (\ref{eq1:cond3}), and
(\ref{eq1:cond4}) are as follows.

\begin{figure}
\includegraphics[width=7cm]{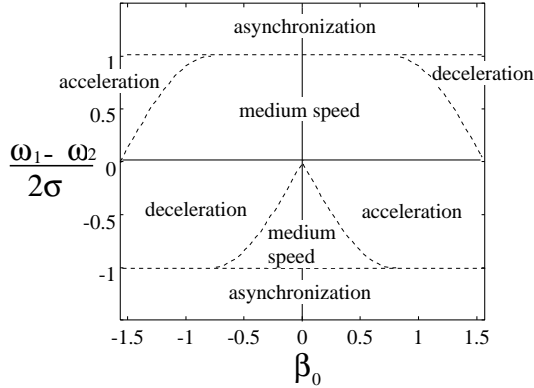}
\caption{Parameter region of acceleration (deceleration) effect of two diffusionally coupled CICs.}
\label{A-C}
\end{figure}

If $\sigma>0$, 
\begin{eqnarray}
& &\sin \eta = - \frac{\omega_1-\omega_2}{2\sigma}, \ \ -\pi/2<\eta<0,
\nonumber \\
& &\left\{ \begin{array}{ll}
-\frac{\pi}{2} <\beta_0 < \frac{-\pi-\eta}{2} & {\rm
(acceleration)} \\
\frac{\pi+\eta}{2} <\beta_0 < \frac{\pi}{2} & {\rm
(deceleration)}
\end{array} \right. .
\end{eqnarray}
If $\sigma<0$, 
\begin{eqnarray}
& &\sin \eta = - \frac{\omega_1-\omega_2}{2\sigma}, \ \ \pi/2<\eta<\pi,
\nonumber \\
& &\left\{ \begin{array}{ll}
\frac{\pi - \eta}{2} <\beta_0 < \frac{\pi}{2} & {\rm
(acceleration)} \\
-\frac{\pi}{2} <\beta_0 < \frac{-\pi+\eta}{2} & {\rm
(deceleration)}
\end{array} \right. .
\end{eqnarray}
Figure \ref{A-C} shows a phase diagram of the acceleration
(deceleration) effect of two diffusionally coupled CICs.

Sakaguchi and Kuramoto proposed a mean field model (the SK model) of
coupled oscillators \cite{kuramoto}. The acceleration (deceleration)
effect is also observed in the SK model from the point of view of our
proposed theory. The SK model is expressed by 
\begin{eqnarray}
\frac{d \phi_i}{d\tau} &=& \omega_i + \frac{J}{N}
\sum_{j (\neq i)}^N \sin(\phi_j - \phi_i+\beta_0), \label{eq.sk}
\end{eqnarray}
where $\phi_i$ is the phase of the $i-$th oscillator (with a total of
$N$ oscillators), and $\omega_i$ represents its natural frequency. The
quantity $J$ represents the strength of the mutual coupling. The
quantity $\beta_0$ in Eq. (\ref{eq.sk}) represents a uniform bias.
Since Eq. (\ref{eq.sk}) can be interpreted as a
system of weakly coupled CICs, $\beta_0$ represents the intersection
angle between the isochron and the orbit, as described above. 
Due to the effect of the bias caused by the
curved isochron sets, the mutual interaction between a pair of
oscillators is asymmetric. Such an unbalanced mutual interaction is the
mechanism of the acceleration (deceleration)
effect. Therefore, we can conclude that the SK model has the minimum
mechanisms for the acceleration (deceleration)
effect related to curved isochron sets. As the SK model is not
frustrated, we need to study frustrated coupled oscillator systems.

\section{Frustrated coupled oscillator systems}

In this section, we extend the SK model to frustrated coupled oscillator
systems with large degrees of freedom and describe the mechanism of the
acceleration (deceleration) effect unique to frustrated coupled
oscillators.

In general, frustrated systems differ from ferromagnetic ones in that
they have the {\it Onsager reaction term} (ORT). Here, we focus on the
effect of the ORT on the acceleration (deceleration) that exists only in
frustrated globally coupled oscillator systems, and in particular cannot
be found in equilibrium systems. To make this effect clear, it would be
best to compare two frustrated systems that, with the exception of a
different quantity of the ORT, have the same order parameter
equations. In addition, these systems should have a clear correspondence
with an equilibrium system because the effects of the ORT are well
understood in equilibrium systems.

We thus consider a system with the following form to be ideal.
\begin{eqnarray}
\frac{d \phi_i}{d\tau} = \omega_i 
 + \sum_{j(\neq i)}^{N} J_{ij}\sin(\phi_j - \phi_i + \beta_{ij} + \beta_0), 
 \label{eq.model}
\end{eqnarray}
This simple phase equation was obtained by approximating
$\Gamma_{ij}(\phi)$ in Eq. (\ref{eq.org}) to the lowest frequency
component. In fact, such systems are commonly used as models of coupled
oscillator systems \cite{kuramoto0,Vicente,daido,bonilla,Izhikevich}.
Natural frequencies $\{ \omega_i \}_{i=1, \cdots, N}$ in
Eq. (\ref{eq.model}) are randomly distributed with a density represented
by $g(\omega)$. Also in Eq. (\ref{eq.model}), $J_{ij}$, and $\beta_{ij}$
denote the amplitude of coupling from unit $j$ to unit $i$ and its
delay, respectively.  In the present study, we have selected the
following two generalized Hebb learning rules with random dilutions
\cite{aoyagi2} to determine $J_{ij}$ and $\beta_{ij}$:
\begin{eqnarray}
K_{ij} &=& J_{ij} \exp(i \beta_{ij}) 
 = \frac{c_{ij}}{c N} \sum_{\mu=1}^{p} 
                 \xi_{i}^{\mu} {\overline\xi}_{j}^{\mu}, \\
 \xi_i^{\mu} &=& \exp( i \theta_i^{\mu}), \\
c_{ij} &=& \left\{ \begin{array}{ll}
1 \quad & {\rm with\  probability}\ c\\
0 \quad & {\rm with\  probability}\ 1-c
\end{array} \right. , 
\end{eqnarray}
where $\overline{\cdot}$ means the complex conjugate. The $\{
\theta_i^\mu \}_{i=1, \cdots, N, \mu=1, \cdots, p}$ are the phase
patterns to be stored in the present model and are assigned to random
numbers with a uniform probability on the interval $[0, 2\pi]$. The
$\mu$ is an index of the stored pattern, and $p$ is the total number of
stored patterns. We define parameter $\alpha$ (the loading rate) by
$\alpha=p/N$. When $\alpha \sim O(1)$, the system is frustrated.  When
$\alpha = 0$, the system is equivalent to the SK model. The quantity
$c_{ij}$ is the dilution coefficient. Let $c_{ij}=1$ if there is
non-zero coupling from unit $j$ to unit $i$ and $c_{ij}=0$
otherwise. The number of fan-ins (fan-outs) is restricted to $O(N)$,
i.e., $c \sim O(1)$.

Here, we consider both symmetric dilution (i.e., $c_{ij}=c_{ji}$) and
asymmetric dilution (i.e., $c_{ij}$ and $c_{ji}$ are independent random
variables)\cite{okada}. The quantity $\beta_0$ in Eq. (\ref{eq.model})
represents a uniform bias. Since Eq. (\ref{eq.model}) can be interpreted
as a system of weakly coupled CICs, $\beta_0$ represents the
intersection angle between the isochron and the orbit, as discussed in
Section III. Due to the effect of the bias caused by the curved isochron
sets, the mutual interaction between a pair of oscillators is
asymmetric, even if $J_{ij}= J_{ji}$ and $\beta_{ij} =-\beta_{ji}$. Such
an unbalanced mutual interaction is the essence of the acceleration
(deceleration) effect, because oscillators are pulled forward from the
isochron, as discussed in Section III. If $\beta_0=0$,
Eq. (\ref{eq.model}) is equivalent to weakly coupled RICs.

There is a close analogy between the phase description of coupled
oscillators and the classical XY-spin model of magnetic material. If
$\omega_i = 0$, $\beta_0 = 0$, $J_{ij}= J_{ji}$, and $\beta_{ij} = -
\beta_{ji}$, we can define the following Lyapnov function on the system:
\begin{eqnarray}
E = - \frac{1}{2} \sum_{i, j}^N J_{i j} \cos(\phi_j - \phi_i +
\beta_{ij}). \label{eq.poten}
\end{eqnarray}
This function enables us to use conventional statistical mechanics for XY-spin systems 
\cite{cook} to analyze coupled oscillators in the equilibrium state. Thus, this system can 
be mapped to an XY-spin system \cite{aoyagi2,cook}. In this way, we can make 
a bridge between the frustrated coupled oscillator system and the equilibrium system.

\section{Order parameter equations}

Let us consider steady states of the system in the limit $\tau
\rightarrow \infty$.  Our theory is based on the condition that there is
one large cluster of oscillators synchronously locked at frequency
$\Omega$ and the number of this cluster scales as $\sim$ $O(N)$. Under
this condition, Daido demonstrated, through a scaling plot obtained from
numerical simulation, that variation in the parameter order scales as
$O(1/\sqrt{N})$ in ferromagnetic systems with one large synchronous
cluster \cite{daido2}. We thus assume that the self-averaging property
holds in our system and that the order parameters are constant in the
limit $N \rightarrow \infty$. These assumptions were also used by
Sakaguchi and Kuramoto \cite{kuramoto}.

Redefining  $\phi_i$ according to $\phi_i \rightarrow  \phi_i + \Omega
\tau$ and substituting this into Eq. (\ref{eq.model}), we obtain 
\begin{eqnarray}
- \frac{d \phi_i}{d\tau} + \omega_i - \Omega = \sin(\phi_i) h^R_i - \cos(\phi_i) 
h^I_i \label{eq:hei1}.
\end{eqnarray}
where $h_i$ represents the so-called "local field", which is described as
\begin{eqnarray}
& &\hspace{-0.3cm}h_i = h_i^R +i h_i^I = e^{i \beta_0} \sum_{j(\neq i)}^{N} 
K_{ij} s_j =e^{i \beta_0}
\nonumber \\
& &\hspace{-0.2cm}\times
\left( \sum_{\mu}^{p} \xi_i^\mu m^\mu + \frac{1}{N}\sum_{\mu}^{p}
\sum_{j(\neq i)}^{N} 
\frac{c_{ij}-c}{c} \xi_{i}^{\mu} {\overline\xi}_{j}^{\mu} s_j -
\alpha s_i \right), \label{eq:lf1} 
\end{eqnarray}
For convenience, we write $s_i = \exp(i \phi_i)$. The order parameter
$m^\mu$, which is the overlap between the system state $\{ s_i \}_{i=1,
\cdots, N}$ and embedded pattern $\{ \xi_i^\mu \}_{i=1, \cdots, N}$, is
defined as 
\begin{eqnarray}
m^\mu = \frac{1}{N} \sum_{j=1}^{N} {\overline\xi}_j^\mu s_j.
\label{eq:ov1}
\end{eqnarray}
In the thermodynamic limit, the effect of the second term of 
Eq. (\ref{eq:lf1}), i.e. 
$\frac{1}{N}\sum_{\mu}^{p} \sum_{j\neq i}^{N} 
\frac{c_{ij}-c}{c} \xi_{i}^{\mu} {\overline\xi}_{j}^{\mu} s_j$,
is equivalent to that of the effect of additive coupling noise 
\cite{Sompolinsky,aoyagi2,okada}:
\begin{eqnarray}
& &K_{ij} = J_{ij} \exp(i \beta_{ij}) 
  = \frac{1}{N} \sum_{\mu=1}^{p}
                 \xi_{i}^{\mu} {\overline\xi}_{j}^{\mu} + \delta
n_{ij}\\
& &{\rm Re} \delta n_{ij} \sim {\cal N}(0, \nu^2/N), \\
& &{\rm Im} \delta n_{ij} \sim {\cal N}(0, \nu^2/N), \\
& &\nu^2 = \frac{\alpha(1-c)}{2c}, \label{eq.val}
\end{eqnarray}
where $\nu^2/N$ is the variance of additive coupling noise $\delta n_{ij}$.
In the case of symmetric dilution, $\delta n_{ij}$ is symmetric, i.e.,
$\delta n_{ij} = \overline{\delta n_{ji}}$.
On the other hand, in the case of asymmetric dilution, 
$\delta n_{ij}$ and $\delta n_{ji}$ are independent random variables. 
In the limit of strong dilution, i.e. $c \rightarrow 0$, with $\alpha /c$ kept finite, our 
system is reduced to a glass oscillator, which corresponds to the Sherrington-Kirkpatrick 
model of spin glass \cite{sherrington}. Therefore, our theory can cover two types of 
frustrated systems, the oscillator associative memory system and the glass oscillator 
system.

In general, the fields $h^R_i$ and $h^I_i$ involve the ORT corresponding to the 
effective self-feedback \cite{fukai3}. We must eliminate the ORT from these fields. 
Here, we assume that the local field splits into a "pure" effective local field, $\tilde{h}_i 
= \tilde{h}^R_i + i \tilde{h}^I_i$, and the ORT, $\Gamma s_i$:
\begin{eqnarray}
h_i= \tilde{h}_i +  \Gamma s_i. \label{eq:as2} 
\end{eqnarray}
We neglect the complex conjugate term of the ORT, which leads to a higher-harmonic 
term of the coupling function \cite{aonishi}. This can be done in the present model 
because we use generalized Hebb learning rules (see Appendix). Hence, by substituting 
Eq. (\ref{eq:as2}) into Eq. (\ref{eq:hei1}), we obtain
\begin{eqnarray}
- \frac{d \phi_i}{dt} + 
\omega_i - \tilde{\Omega} = \sin(\phi_i) \tilde{h}^R_i -
\cos(\phi_i) \tilde{h}^I_i,  \label{eq:hei2} \\
\tilde{\Omega} = \Omega - \left|\Gamma \right|\sin(\psi), \ \
\psi = {\rm Arg} \left(\Gamma \right), \label{eq:gamma} 
\end{eqnarray}
which does not contain the ORT.
The quantity $\tilde{\Omega}$ represents the effective frequency of the synchronous 
oscillators. We can regard $\tilde{\Omega}$ as the renormalized version of $\Omega$, 
from which the ORT has been removed, so $\tilde{\Omega}$ takes a different value 
from the observable $\Omega$ in general.
Thus, $\Omega - \tilde{\Omega}$ represents the contribution of the ORT to the 
acceleration (deceleration) effect; 
$\tilde{\Omega}$ is one of the order parameters of our theory. In the analysis that 
follows, $\tilde{h}_i$ and $\Gamma$ are obtained in a self-consistent manner (see 
Appendix). 

Let us consider synchronous oscillators in which
$\frac{d\phi_i}{d\tau}=0$ is satisfied in Eq. (\ref{eq:hei2}).  As a
result, the stationary states of the oscillators are satisfied:
\begin{eqnarray}
s_i = \frac{\tilde{h}_i}{|\tilde{h}_i|}
\frac{i (\omega_i-\tilde{\Omega})  +  \sqrt{|\tilde{h}_i|^2 - 
(\omega_i-\tilde{\Omega})^2}}{|\tilde{h}_i|}.  \label{eq.stationary}
\end{eqnarray}
From Eq. (\ref{eq.stationary}), 
we obtain the following condition for synchronization:
\begin{eqnarray}
|\tilde{h}_i|^2 \geq (\omega_i-\tilde{\Omega})^2. \label{eq.cond}
\end{eqnarray}
If $\tilde{h}_i$ does not satisfy Eq. (\ref{eq.cond}), 
$\phi_i$ continues rotating individually (i.e., desynchronization).
This condition corresponds to
Eq. (\ref{eq1:cond2}) for two coupled oscillators.
We assume that the microscopic memory effect can be neglected in the $\tau 
\rightarrow \infty$ limit. In other words, the asymptotic state of the system as $\tau 
\rightarrow \infty$ is assumed to be independent of the initial conditions at $\tau = 0$. 
This assumption is exact in non-frustrated systems \cite{bonilla}. Under this 
assumption, $\phi_i$ takes the following form:
\begin{eqnarray}
\phi_i = \tilde{\omega}_i \tau +f(\tilde{\omega}_i \tau). \label{eq.as}
\end{eqnarray}
where $f(x)$ is a periodic function with period $2 \pi$, and $\tilde{\omega}_i$ is the 
resultant frequency of asynchronous oscillators into which the ORT has been absorbed.
$\tilde{\omega}_i$  is given by the following equation:
\begin{widetext}
\begin{eqnarray}
\tilde{\omega}_i= \tilde{\Omega} + (\omega_i-\tilde{\Omega}) \sqrt{1 - 
\frac{|\tilde{h}_i|^2}{(\omega_i-\tilde{\Omega})^2}}.
\label{eq:fr}
\end{eqnarray}
Applying the SK theory \cite{kuramoto} to Eq. (\ref{eq:hei2}), we obtain the average 
of $s_i$ over $\omega_i$:
\begin{eqnarray}
\left<s_i \right>_\omega &=& \tilde{h}_i \int_{-\pi/2}^{\pi/2} d \phi 
g\left(\tilde{\Omega} +
|\tilde{h}_i|\sin\phi \right)\cos\phi\exp(i \phi)\nonumber \\ 
&+& i  \tilde{h}_i \int_{0}^{\pi/2} d \phi 
\frac{\cos\phi (1-\cos\phi)}{\sin^3\phi}  
\left\{g\left(\tilde{\Omega}+\frac{|\tilde{h}_i|}{\sin\phi}\right)-g\left(\tilde{\O
mega}-\frac{|\tilde{h}_i|}{\sin\phi}\right)
\right\}. 
\end{eqnarray}

In this analysis, we focus on the memory retrieval states in which the configuration has 
appreciable overlap with the condensed pattern $\mbox{\boldmath $\xi$}^1$ ($m^1 
\sim O(1)$) and has little overlap with the uncondensed patterns $\mbox{\boldmath 
$\xi$}^\mu$ for $\mu>1$ ($m^\mu \sim O(1/\sqrt{N})$). Under this assumption, we 
obtain the contribution of the uncondensed patterns using SCSNA (self-consistent signal 
to noise analysis) \cite{fukai3} and determine $\tilde{h}_i$ in a self-consistent manner 
(see Appendix). Finally, the equations relating the order parameters $|m^1|$, $U$, and 
$\tilde{\Omega}$ are obtained using the self-consistent local field:
\begin{eqnarray}
|m^1| e^{-i\beta_0}&=& \left<\left< \tilde{X}(x_1,x_2;\tilde{\Omega}) 
\right>\right>_{x_1, x_2}, \label{eq.o1} \\
U e^{-i\beta_0}&=& \left<\left< F_1(x_1,x_2;\tilde{\Omega}) \right>\right>_{x_1, 
x_2}, \label{eq.o2}
\end{eqnarray}
where $\left<\left< \cdots \right>\right>_{x_1, x_2}$ is the Gaussian average over 
$x_1$ and $x_2$,
$\left<\left< \cdots \right>\right>_{x_1, x_2} = \int
\int Dx_1 Dx_2\cdots$. The quantity $U$ corresponds to the susceptibility, which is 
the measure of the sensitivity to external fields. Since the present system possesses 
rotational symmetry with respect to the phase $\phi_i$, we can safely set the condensed 
pattern to $\xi^1_i=1$.
Now, $\tilde{h}$, $\tilde{X}$, $F_1$, and $Dx_1 Dx_2$ can be expressed as
\begin{eqnarray}
& &Dx_1 Dx_2=\frac{dx_1 dx_2}{2 \pi \rho^2} 
\exp\left(-\frac{x_1^2+x_2^2}{2 \rho^2} \right),\\
& &\rho^2 = \frac{\alpha}{2|1-U|^2} + \nu^2, \ \
\nu^2 = \frac{\alpha(1-c)}{2c}, 
\hspace{0.3cm} \tilde{h} = |m^1| + x_1+i x_2, \\
\tilde{X}(x_1,x_2;\tilde{\Omega}) &=& \tilde{h} \int_{-\pi/2}^{\pi/2} d \phi 
g\left(\tilde{\Omega} +
|\tilde{h}|\sin\phi \right)\cos\phi\exp(i \phi)\nonumber \\ 
& &+ i  \tilde{h} \int_{0}^{\pi/2} d \phi 
\frac{\cos\phi (1-\cos\phi)}{\sin^3\phi}  
\left\{ g\left(\tilde{\Omega}+\frac{|\tilde{h}|}{\sin\phi}\right)-g\left(\tilde{\Omega}-\frac{|\tilde{h}|}{\sin\phi}\right) \right\}, \label{eq:aa} \\
F_1(x_1,x_2;\tilde{\Omega}) &=&
\int_{-\pi/2}^{\pi/2} d \phi \left( g\left(\tilde{\Omega}+|\tilde{h}|\sin\phi \right)
 + \frac{|\tilde{h}|}{2} \sin\phi
g'\left(\tilde{\Omega}+|\tilde{h}|\sin\phi \right)\right)  
\cos\phi\exp(i \phi)\nonumber \\ 
& &+ i  \int_{0}^{\pi/2} d \phi 
\frac{\cos\phi (1-\cos\phi)}{\sin^3\phi} 
\left\{ 
g\left(\tilde{\Omega}+\frac{|\tilde{h}|}{\sin\phi}\right)-g\left(\tilde{\Omega}-\frac{|\tilde{h}|}{\sin\phi}\right)
\right\} \nonumber \\ 
& &+ i  \frac{|\tilde{h}|}{2}\int_{0}^{\pi/2} d \phi 
\frac{\cos\phi (1-\cos\phi)}{\sin^4\phi}  
\left\{ 
g'\left(\tilde{\Omega}+\frac{|\tilde{h}|}{\sin\phi}\right)+g'\left(\tilde{\Omega}-
\frac{|\tilde{h}|}{\sin\phi}\right) \right\} .\label{eq:bb}
\end{eqnarray}
The terms with the coefficient $i$ in Eqs. (\ref{eq:aa}) and (\ref{eq:bb}) represent the 
contribution of asynchronous oscillators to the macroscopic
 behavior. The other terms represent the contribution of the cluster of synchronous oscillators. In the case of the 
symmetric diluted system, $\Gamma$ can be expressed as 
\begin{eqnarray}
\Gamma e^{-i\beta_0} = \frac{\alpha U}{1-U} + \frac{\alpha (1-c)}{c} U. 
\label{eq:ga1}
\end{eqnarray}
In the case of the asymmetric diluted system, on the other hand, we have
\begin{eqnarray}
\Gamma e^{-i\beta_0} = \frac{\alpha U}{1-U}. \label{eq:ga2}
\end{eqnarray}

$\tilde{h}$ and $\tilde{\Omega}$ are the renormalized versions of $h$ and 
$\Omega$, respectively, from which the ORT has been removed, and thus 
$\tilde{h}$ and $\tilde{\Omega}$ are independent of the ORT. Therefore, the two 
models we consider have identical order parameter equations, (\ref{eq.o1}) and 
(\ref{eq.o2}), written  using the term of the renormalized quantities $\tilde{h}$ and 
$\tilde{\Omega}$.
From Eq. (\ref{eq:gamma}), the difference between the ORTs in Eqs. (\ref{eq:ga1}) 
and (\ref{eq:ga2}) leads to a different value for the observable $\Omega$ only when 
$\beta_0 \neq 0$. In this way we are able to clearly separate the effect of the ORT, and 
therefore, by observing the macroscopic parameter $\Omega$ of these two systems, we 
can analyze the effect of the ORT qualitatively and quantitatively. 

The distribution of renormalized resultant frequencies $\tilde{\omega}$ (Eq. 
(\ref{eq:fr})) in the memory retrieval state, which is denoted as
$\tilde{p}(\tilde{\omega})$, becomes 
\begin{eqnarray}
\tilde{p}(\tilde{\omega}) &=& r
\delta(\tilde{\omega}-\tilde{\Omega}) +
\int Dx_1 Dx_2\frac{g\left( \tilde{\Omega}+ (\tilde{\omega}- 
\tilde{\Omega})\sqrt{1 + 
\frac{|\tilde{h}|^2}{(\tilde{\omega}-\tilde{\Omega})^2}}\right)}{
\sqrt{1 +
\frac{|\tilde{h}|^2}{(\tilde{\omega}-\tilde{\Omega})^2}}}, \label{eq:dis2} \\
r &=&\int Dx_1 Dx_2 |\tilde{h}| \int_{-\pi/2}^{\pi/2} d\phi
g\left(\tilde{\Omega} + |\tilde{h}| \sin\phi \right) \cos\phi ,
\end{eqnarray}
The $\delta$-function in Eq. (\ref{eq:dis}) indicates the cluster of oscillators 
synchronously locked at frequency $\tilde{\Omega}$. The value $r$ is the ratio 
between the number of synchronous oscillators and the total number of oscillators $N$.
The second term in Eq. (\ref{eq:dis}) represents the distribution of asynchronous 
oscillators. From the distribution given by Eq. (\ref{eq:dis2}), the distribution of 
observable resultant frequencies $\overline{\omega}$, which is denoted as 
$p(\overline{\omega})$, becomes 
\begin{eqnarray}
p(\overline{\omega}) = \tilde{p}\left( \overline{\omega} - (\Omega - 
\tilde{\Omega})\right). \label{eq:dis}
\end{eqnarray}

We now consider the relationships between the present theory and previously proposed 
theories. If $\beta_0=0$ and $g(\omega)$ are symmetric, our theory reduces to the 
theory proposed by Aonishi et al. \cite{aonishi2}:
\begin{eqnarray}
& &\tilde{X}(x_1,x_2;\tilde{\Omega}) = \tilde{h} \int_{-1}^{1} dx  
g\left(|\tilde{h}| x \right) \sqrt{1-x^2},\\
& &F_1(x_1,x_2;\tilde{\Omega})= \int_{-1}^{1}dx 
\left( g \left(|\tilde{h}| x \right)+ \frac{|\tilde{h}|}{2} x g' \left(|\tilde{h}| x \right) 
\right)
\sqrt{1-x^2}.
\end{eqnarray}
where $\tilde{\Omega}=\Omega=0$, since $F_1$ and $U$ are real numbers. 
If $g(\omega)=\delta(\omega)$, $\beta_0 = 0$, and
$c_{ij}=c_{ji}$, where the present model reduces to 
an XY-spin system, we obtain
\begin{eqnarray}
\tilde{X} = \frac{\tilde{h}}{|\tilde{h}|},\hspace{0.2cm}
F_1 = \frac{1}{2|\tilde{h}|},\hspace{0.2cm} 
p(\overline{\omega}) = \delta(\overline{\omega}),
\end{eqnarray}
which coincide with the replica theory of Cook \cite{cook} and SCSNA
\cite{aoyagi3}.
In addition, in the uniform-system limit, $\alpha \rightarrow 0$, our theory 
reproduces the SK theory \cite{kuramoto}:
\begin{eqnarray}
|m^1| e^{-i\beta_0} &=& |m^1| \int_{-\pi/2}^{\pi/2} d \phi g\left(\tilde{\Omega} +
|m^1|\sin\phi \right)\cos\phi\exp(i \phi)\nonumber \\ 
&+& i  |m^1| \int_{0}^{\pi/2} d \phi 
\frac{\cos\phi (1-\cos\phi)}{\sin^3\phi} \left\{ 
g\left(\tilde{\Omega}+\frac{|m^1|}{\sin\phi}\right)-g\left(\tilde{\Omega}-\frac{|
m^1|}{\sin\phi}\right)
\right\}, \\
p(\overline{\omega}) &=& r
\delta(\overline{\omega}-\Omega)+
\frac{g\left( \Omega+ (\overline{\omega}- \Omega)\sqrt{1 + 
\frac{|m^1|^2}{(\overline{\omega}-\Omega)^2}}\right)}{
\sqrt{1 + \frac{|m^1|^2}{(\overline{\omega}-\Omega)^2}}}, \\
r &=& |m^1| \int_{-\pi/2}^{\pi/2} d \phi g\left(\Omega +
|m^1|\sin\phi \right)\cos\phi, \\
\tilde{\Omega} &=& \Omega.
\end{eqnarray}

In the limit $N \rightarrow \infty$, models with random dilution are equivalent to 
models with additive coupling noise \cite{Sompolinsky,aoyagi2,okada}.
In the limit $c \rightarrow 0$ with $\alpha /c$ kept finite, our system is equivalent to a 
glass oscillator system with complex interaction. The equations relating the order 
parameters $|m^1|$ and $\tilde{\Omega}$ are
\begin{eqnarray}
|m^1| e^{-i\beta_0} &=& \left<\left< \tilde{X}(x_1,x_2;\tilde{\Omega})
\right>\right>_{x_1, x_2}, \label{eq:og}
\end{eqnarray}
\begin{eqnarray}
& &Dx_1 Dx_2=\frac{dx_1 dx_2}{2 \pi \nu^2} 
\exp\left(-\frac{x_1^2+x_2^2}{2 \rho^2} \right),\\
& &\rho^2=\nu^2 = \frac{\alpha}{2c}, \hspace{0.3cm} \tilde{h}
= |m^1| + x_1+i x_2, \\
& &\tilde{X}(x_1,x_2;\tilde{\Omega}) = \tilde{h} \int_{-\pi/2}^{\pi/2} d \phi 
g\left(\tilde{\Omega} +
|\tilde{h}|\sin\phi \right)\cos\phi\exp(i \phi)\nonumber \\ 
& &+ i  \tilde{h} \int_{0}^{\pi/2} d \phi 
\frac{\cos\phi (1-\cos\phi)}{\sin^3\phi} \left\{ 
g\left(\tilde{\Omega}+\frac{|\tilde{h}|}{\sin\phi}\right)-g\left(\tilde{\Omega}-\frac{|\tilde{h}|}{\sin\phi}\right) \right\} .
\end{eqnarray}
\end{widetext}
The above order parameter equations do not contain susceptibility $U$.

\section{Simulation}

\subsection{Acceleration (deceleration) effect}

In the numerical simulations we now discuss, we set the distribution of
natural frequencies as 
\begin{eqnarray}
g(\omega)=(2\pi\sigma^2)^{-1/2} \exp(-\omega^2/2\sigma^2), \label{eq:gauss}
\end{eqnarray}
and we used the Euler scheme with a time increment of 0.1, which gave a
sufficiently good approximation compared to that of smaller time
increments. The resultant frequencies $\overline{\omega}_i$ were
calculated using a long time average of $d \phi_i / d \tau$. The
acceleration (deceleration) effect was defined as folows: one large
cluster of oscillators are synchronously locked at a faster (slower)
speed than the mean natural rotation speed of all the oscillators. 

\begin{figure}
(a)\includegraphics[width=6cm]{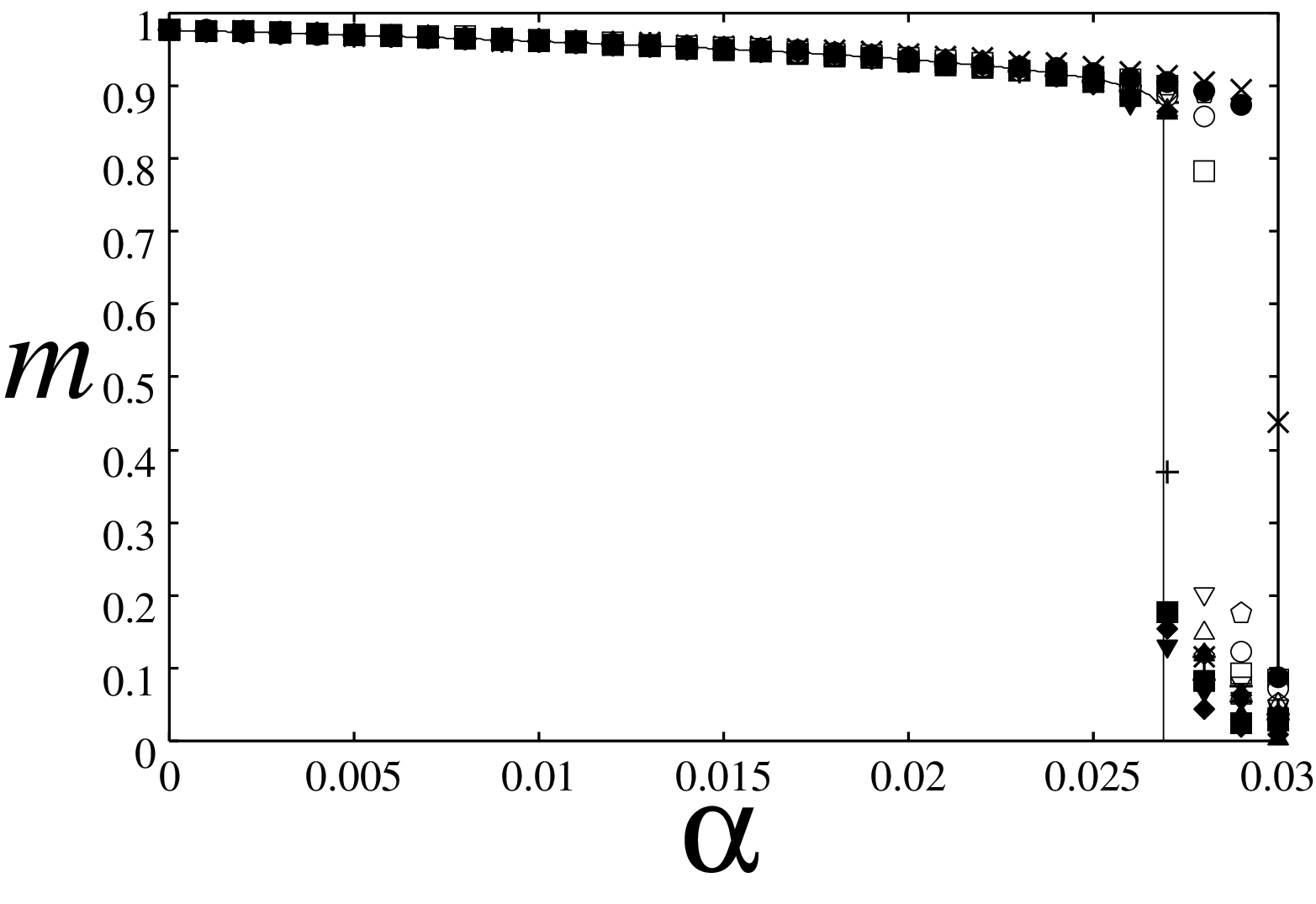}
(b)\includegraphics[width=6cm]{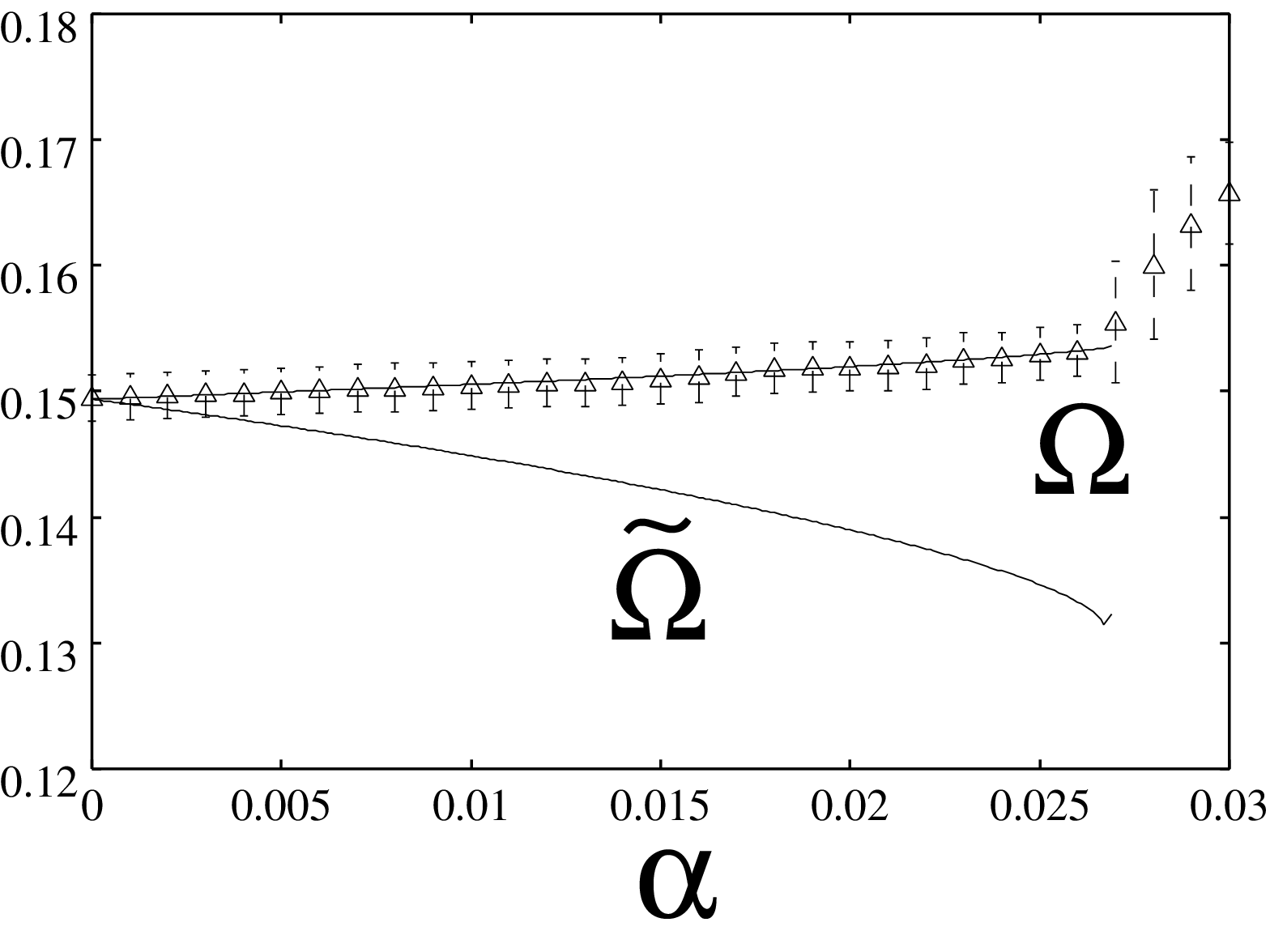}\\
(c)\includegraphics[width=6cm]{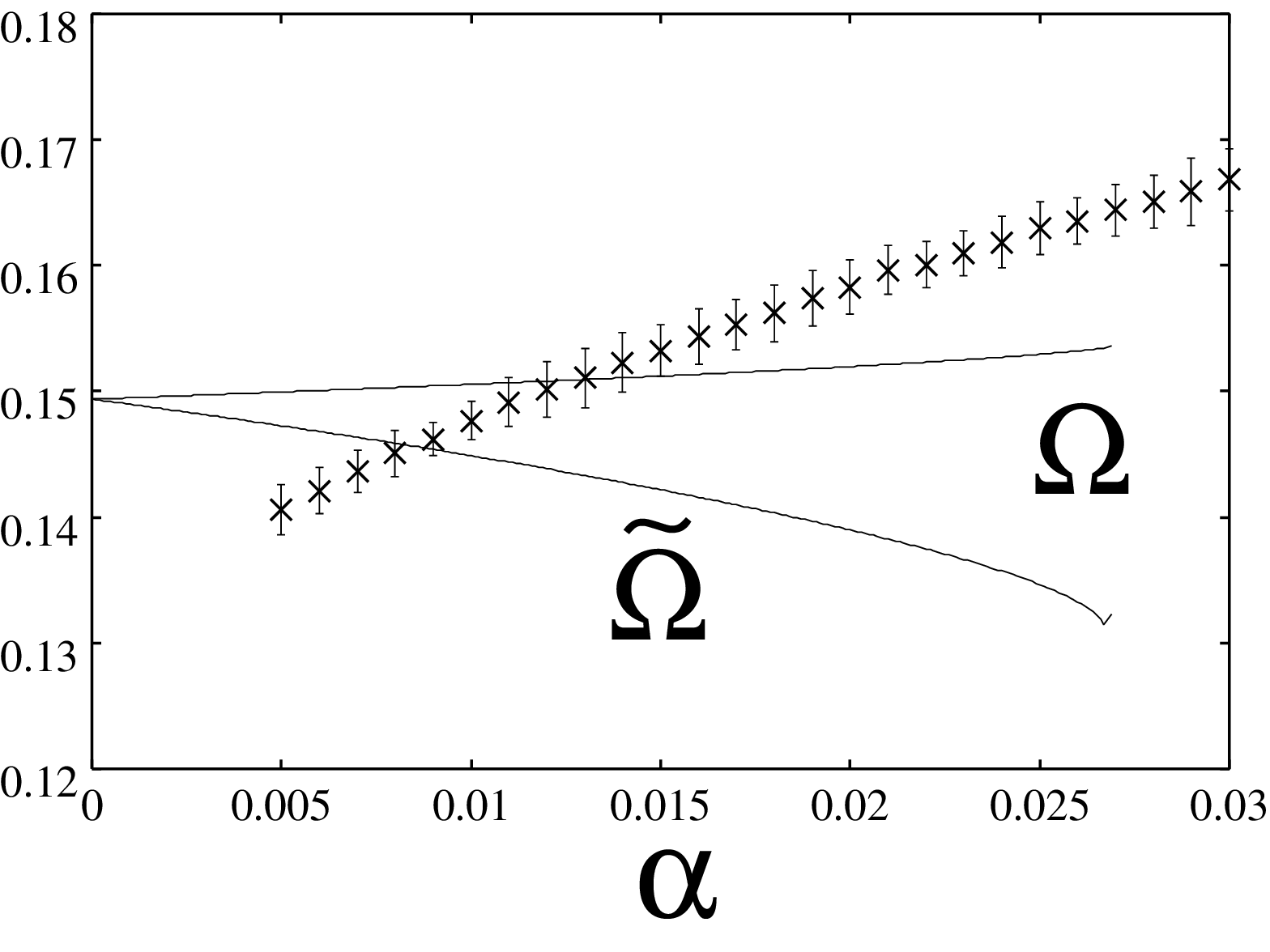}
(d)\includegraphics[width=6cm]{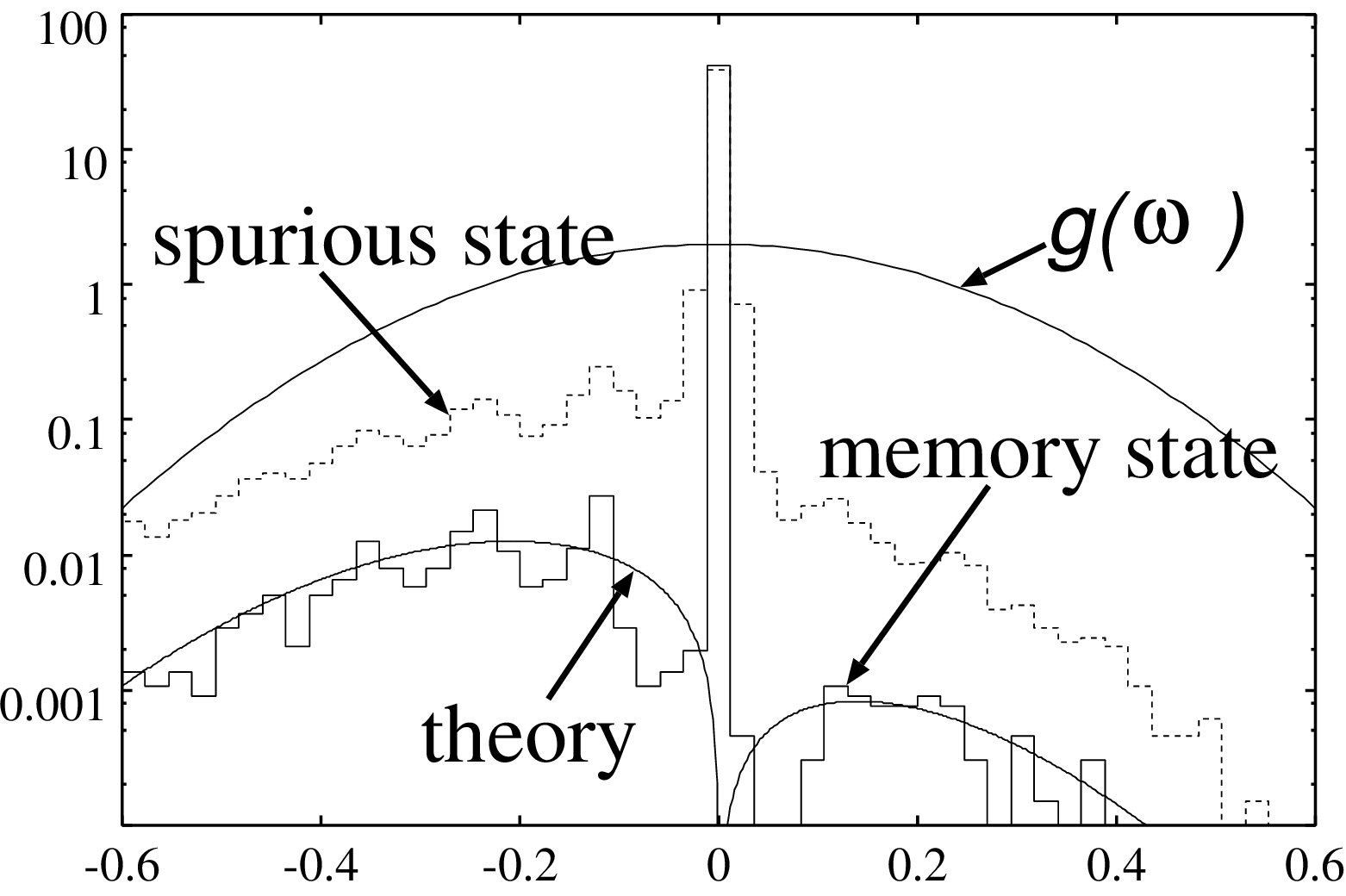}
\caption{Simulated and theoretical results when \mbox{$\beta_0=\pi/20$}, 
\mbox{$\sigma=0.2$}, \mbox{$c=1.0$}, \mbox{$N=20000$}. Solid curves were 
theoretically obtained; plots were obtained by numerical simulation. (a) 
\mbox{$|m^1|$} as a function of \mbox{$\alpha$}. (b) \mbox{$\Omega$} as a 
function of \mbox{$\alpha$} in  memory states. (c) 
\mbox{$\Omega$} as a function of \mbox{$\alpha$} in spurious memory 
states. To compare \mbox{$\Omega$} in spurious memory 
states with that in the memory states, we superimposed theoretical \mbox{$\Omega$} in the 
memory states. (d) Distribution of resultant frequencies 
\mbox{$\overline{\omega}_i$} in memory and spurious memory states. 
\mbox{$\alpha = 0.022$}. The center of the delta peak shifted to \mbox{$0$}.}
\label{tho2}
\end{figure}

First, we set $\beta_0=\pi/20$, $c=1.0$ (i.e., no dilution), and
$\sigma=0.2$. Figure \ref{tho2}(a) shows $|m^1|$ as a function of
$\alpha$, and Fig. \ref{tho2}(b) shows $\Omega$ and $\tilde{\Omega}$ as
functions of  $\alpha$ in the memory states. The solid curves
were obtained theoretically, and the data points with error bars
represent results obtained by numerical simulation. As previously
discussed, the ORT was removed from $\tilde{\Omega}$, so the value of
$\tilde{\Omega}$ differed from that of the observable $\Omega$. The gap
between $\Omega$ and $\tilde{\Omega}$ in Figure \ref{tho2}(b) is in
proportion to the absolute value of the ORT, as described in
Eq. (\ref{eq:gamma}). Thus, increasing loading rate $\alpha$ tended to
accelerate all of the oscillators due to the effect of the ORT. Figure
\ref{tho2}(c) shows $\Omega$ as a function of $\alpha$ in the spurious
memory states, where $\Omega$ corresponds to the maximum point
of a histogram of  $\overline{\omega}_i$. Compared to
Fig. \ref{tho2}(b), the profile of the curve in Fig. \ref{tho2}(c) is
different from that of the memory states.  This
is a very important phenomenon in the context of engineering
because, from the results shown in Figs. \ref{tho2}(b) and (c), we can
determine if the recall process is successful or not by estimating  
the difference in  the rotation speeds of the oscillators. Figure
\ref{tho2}(d) shows histograms of the resultant frequencies
$\overline{\omega}_i$ in the memory and spurious memory
states, which were obtained by numerical simulation ($\alpha=0.022$). In
this graph, we shifted the center of the peak to $0$ and superimposed
the solid curves obtained theoretically for the memory states. 
The theoretical memory-state results are in good agreement with the
simulation ones.

\begin{figure}
(a)\includegraphics[width=6cm]{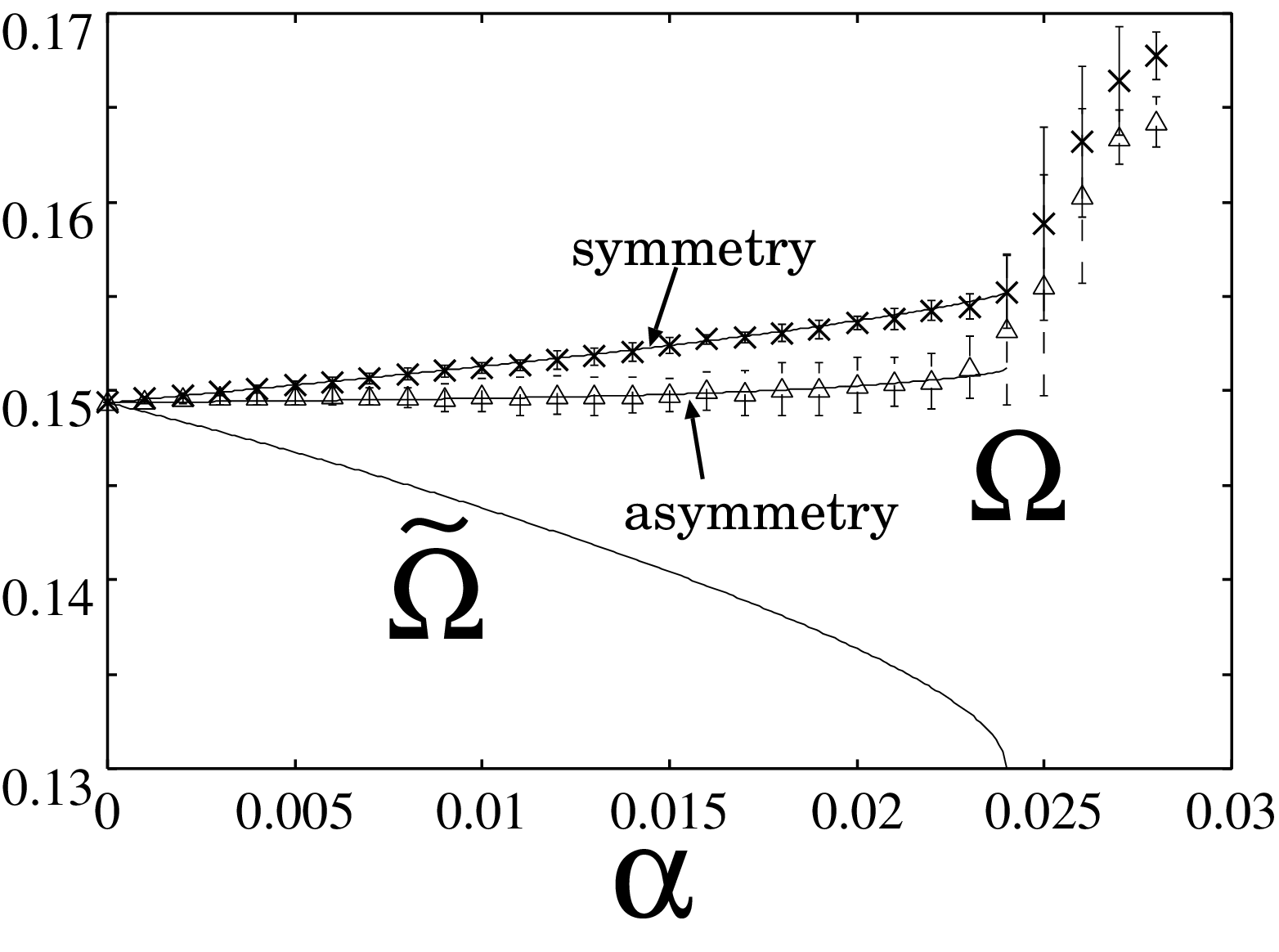}\\
(b)\includegraphics[width=6cm]{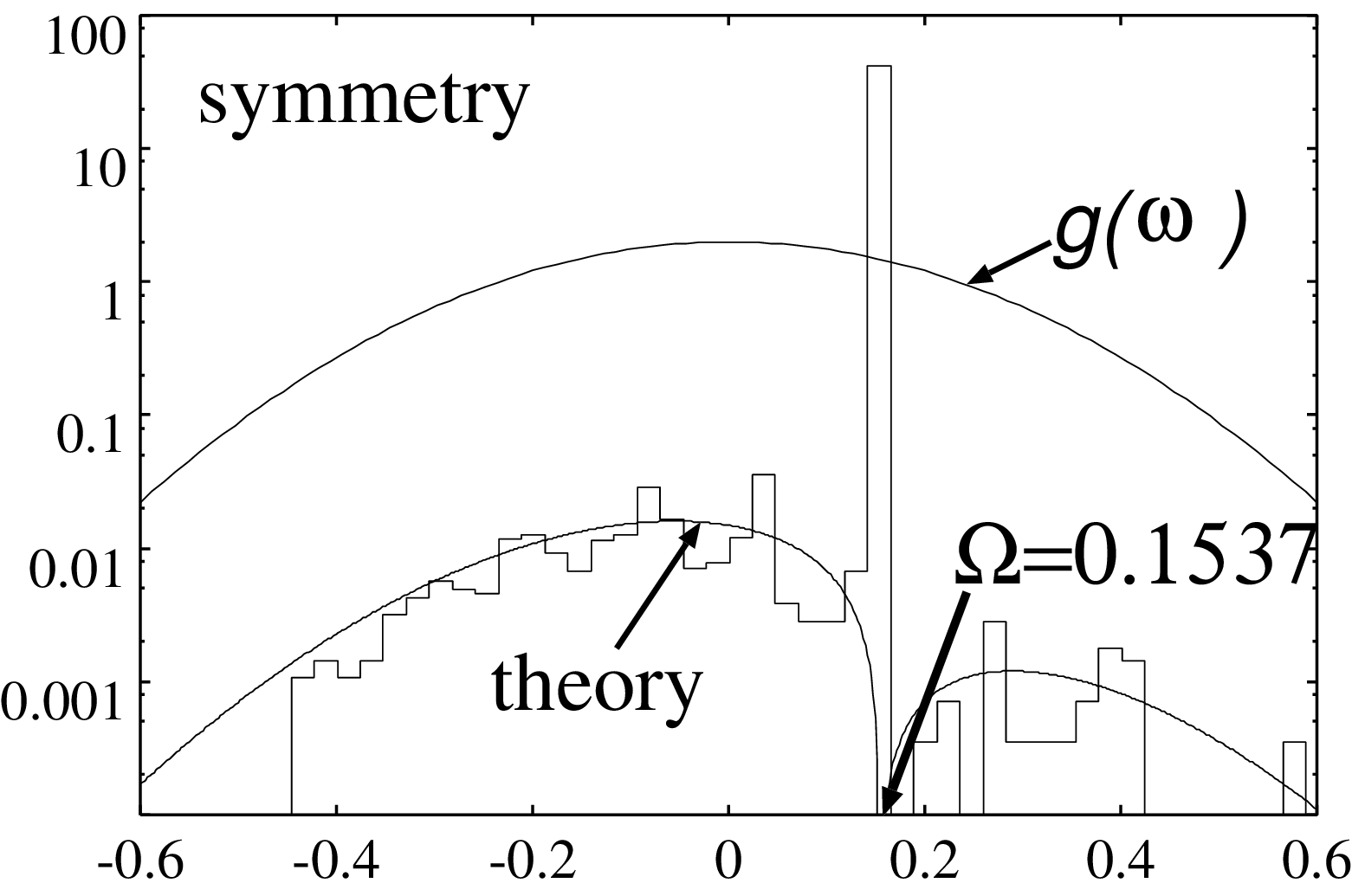}
(c)\includegraphics[width=6cm]{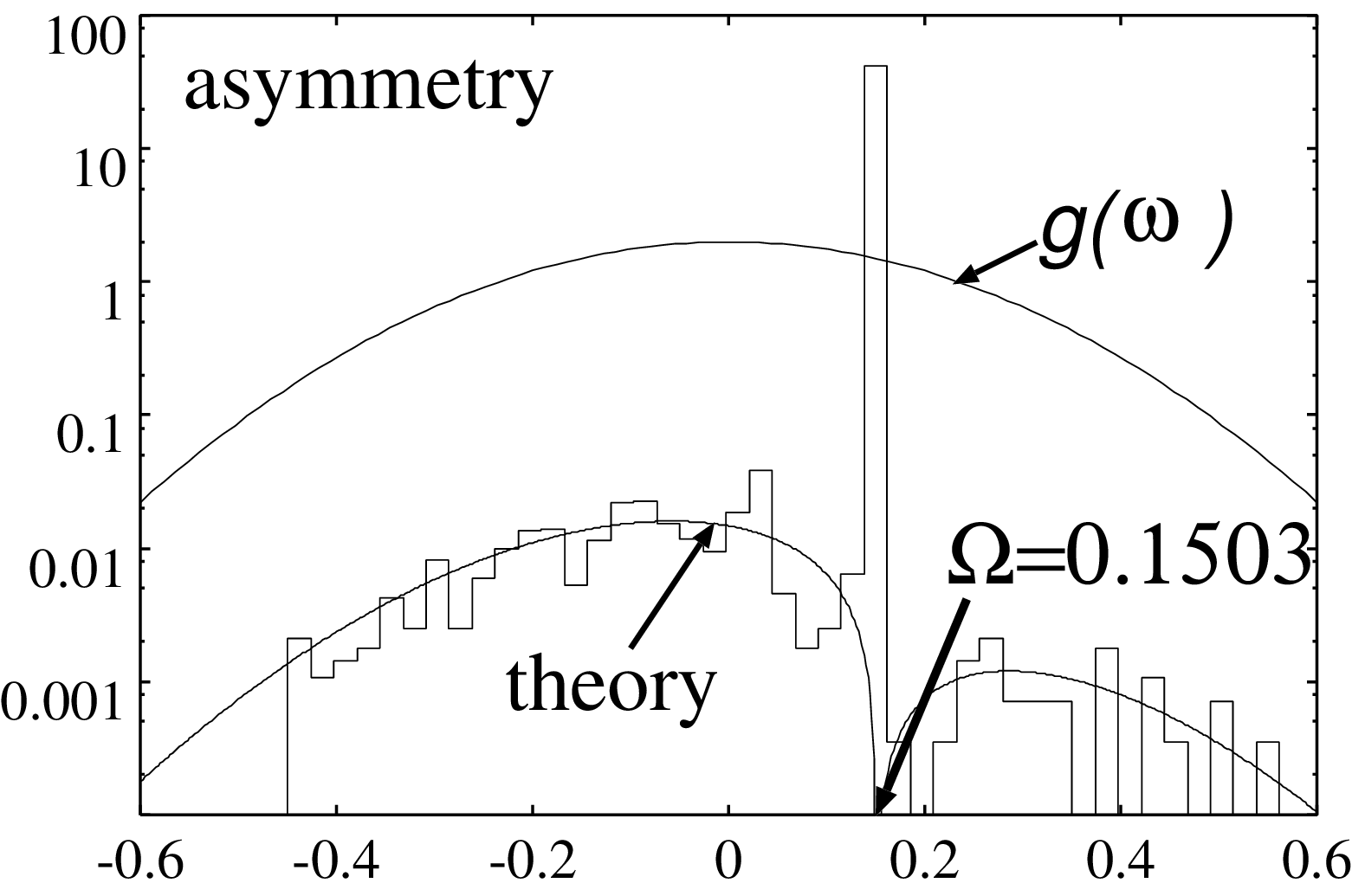}
\caption{Difference between symmetric and asymmetric dilution
systems; \mbox{$N=10,000$}, \mbox{$\sigma=0.2$}, \mbox{$\beta_0=\pi/20$}, 
and \mbox{$c=0.5$}.
(a) $\Omega$ as a function of \mbox{$\alpha$}.
(b) Distribution of resultant frequencies for symmetric dilution system 
(\mbox{$\alpha=0.02$}).
(c) Distribution of resultant frequencies for asymmetric dilution system 
(\mbox{$\alpha=0.02$}).}
\label{tho3}
\end{figure}

Next, to make confirm that the acceleration (deceleration) effect is
caused by the ORT, we analyzed oscillator associative memory models 
involving two types of diluted couplings.  We
set $\sigma=0.2$, $\beta_0=\pi/20$, and $c=0.5$. Figure \ref{tho3}(a)
shows $\Omega$ as a function of  $\alpha$ in the memory retrieval
states; the solid curves were obtained theoretically, and the
data points with error bars represent results obtained by numerical
simulation. It shows that the oscillator rotated faster in the symmetric
diluted system than in the asymmetric one.  As previously
discussed, $\tilde{\Omega}$ in Fig. \ref{tho3}(a),
which represents the effective frequency of synchronous oscillators,
does not depend on the type of dilution, while the observed $\Omega$
strongly depended on it. This dependence was due to the existence of the
ORT. If local field $h$ does not contain the ORT \cite{yoshioka}, plots
obtained from numerical simulations of both models should fit the curve
of $\tilde{\Omega}$. Therefore, the dependence of the observed $\Omega$
on the type of dilution is strong evidence for the existence of the ORT
in the present system. In this figure, we shifted the numerical values
of $\Omega$ at $\alpha=0$ (in the computer simulation) to their
corresponding theoretical values at $\alpha=0$ in order to cancel
fluctuations in the mean value of $g(\omega)$ caused by the finite-size
effect.

\begin{figure}
\begin{center}
(a)\includegraphics[width=6cm]{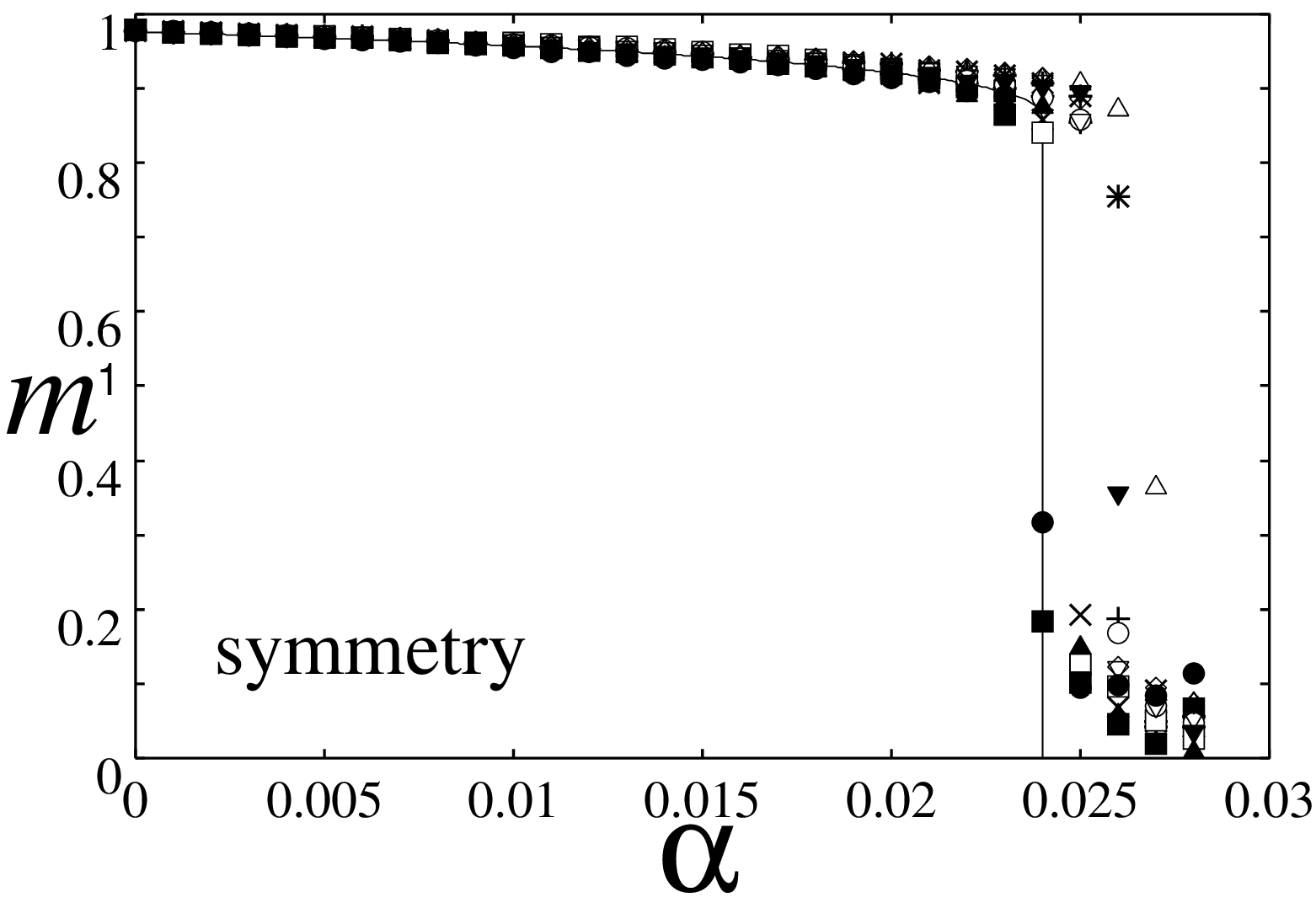}
(b)\includegraphics[width=6cm]{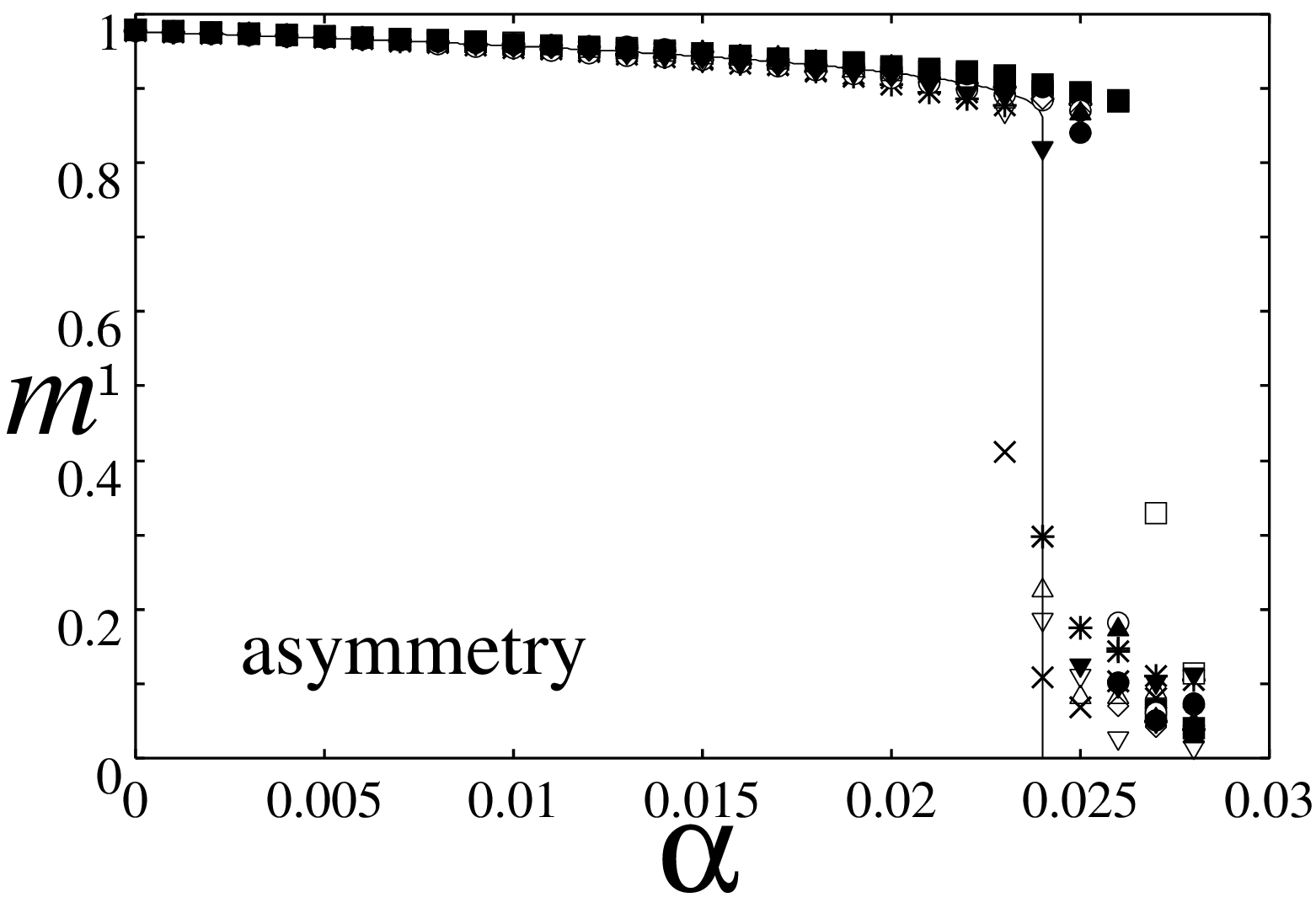}
\end{center}
\caption{\mbox{$|m^1|$} as a function of \mbox{$\alpha$}. Solid curves were 
obtained theoretically, and data points were obtained by numerical simulation; 
\mbox{$N=10,000$}, \mbox{$\sigma=0.2$}, \mbox{$\beta_0=\pi/20$}, and 
\mbox{$c=0.5$}.
(a) Symmetric dilution system. (b) Asymmetric dilution system.}
\label{overlap3}
\end{figure}

Figures \ref{tho3}(b) and (c) show the distributions of the resultant
frequencies for the symmetric and asymmetric dilution systems,
respectively. The theoretical results (solid curve) are in good
agreement with the simulated one (histogram). From the results given in
Figs. \ref{tho3}(b) and (c), the distribution of the
resultant frequencies for the symmetric diluted system is 
identical to that for the asymmetric diluted system, except for
differences in positions caused by the ORT.  We thus conclude that the
mean field, $\tilde{h}$, of the symmetric diluted system is identical to
that of the asymmetric diluted system, since $\tilde{h}$ reflects the
distribution of resultant frequencies, as represented by
Eq. (\ref{eq:dis}). Figures \ref{overlap3}(a) and (b) show $|m^1|$ as a
function of $\alpha$ for the symmetric and asymmetric diluted systems,
respectively. The solid curves were obtained theoretically, and the data
plots represent the results obtained from numerical simulations. As the
figures show, the critical memory capacities of the two models are equal.

Consequently, symmetric and asymmetric diluted systems have the same
macroscopic properties, with the exception of the acceleration
(deceleration) effect caused by the ORT. The quantity of the
ORT depends on the type of dilution, and this dependence leads to a
difference in the rotation speeds of the oscillators for the two cases,
as shown in Fig. \ref{tho3}(a).

\subsection{Glass oscillators}

\begin{figure}
\includegraphics[width=8cm]{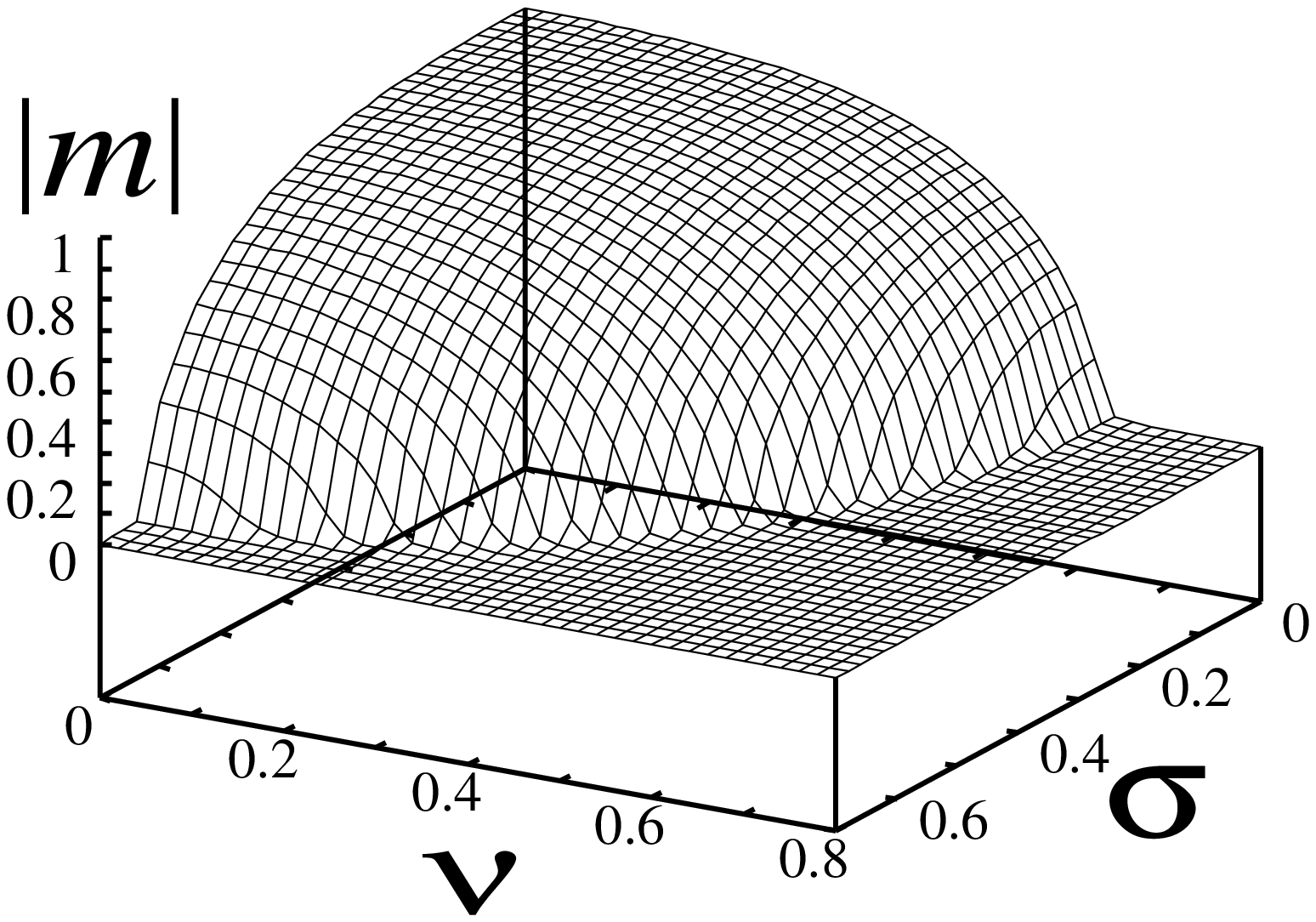}
\caption{Phase diagram in \mbox{$(|m^1|, \sigma, \nu)$} space 
(\mbox{$\beta_0= 0$}) of spin glass model.}
\label{phasediagram2}
\end{figure}

\begin{figure}
\includegraphics[width=4cm]{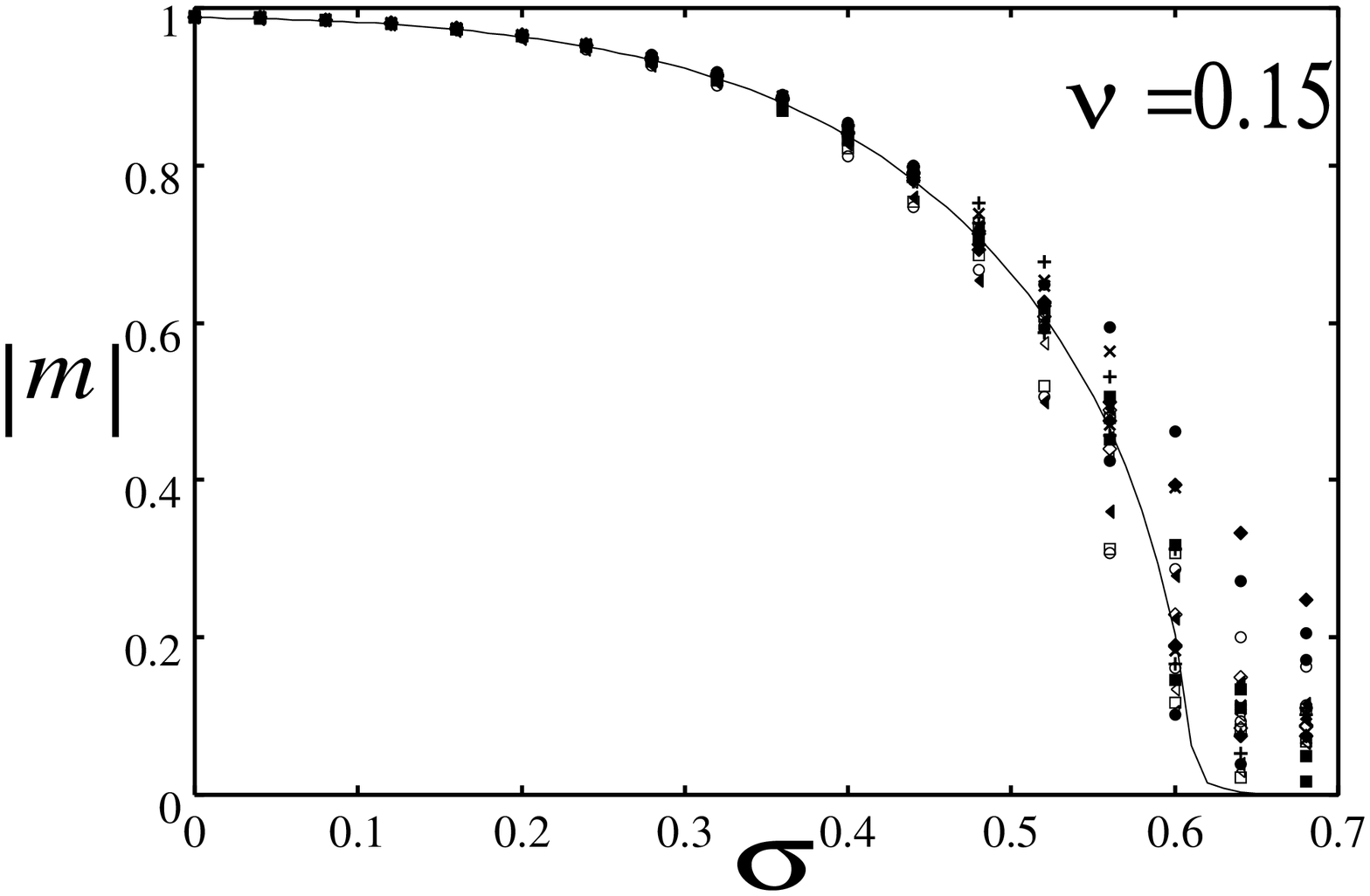}
\includegraphics[width=4cm]{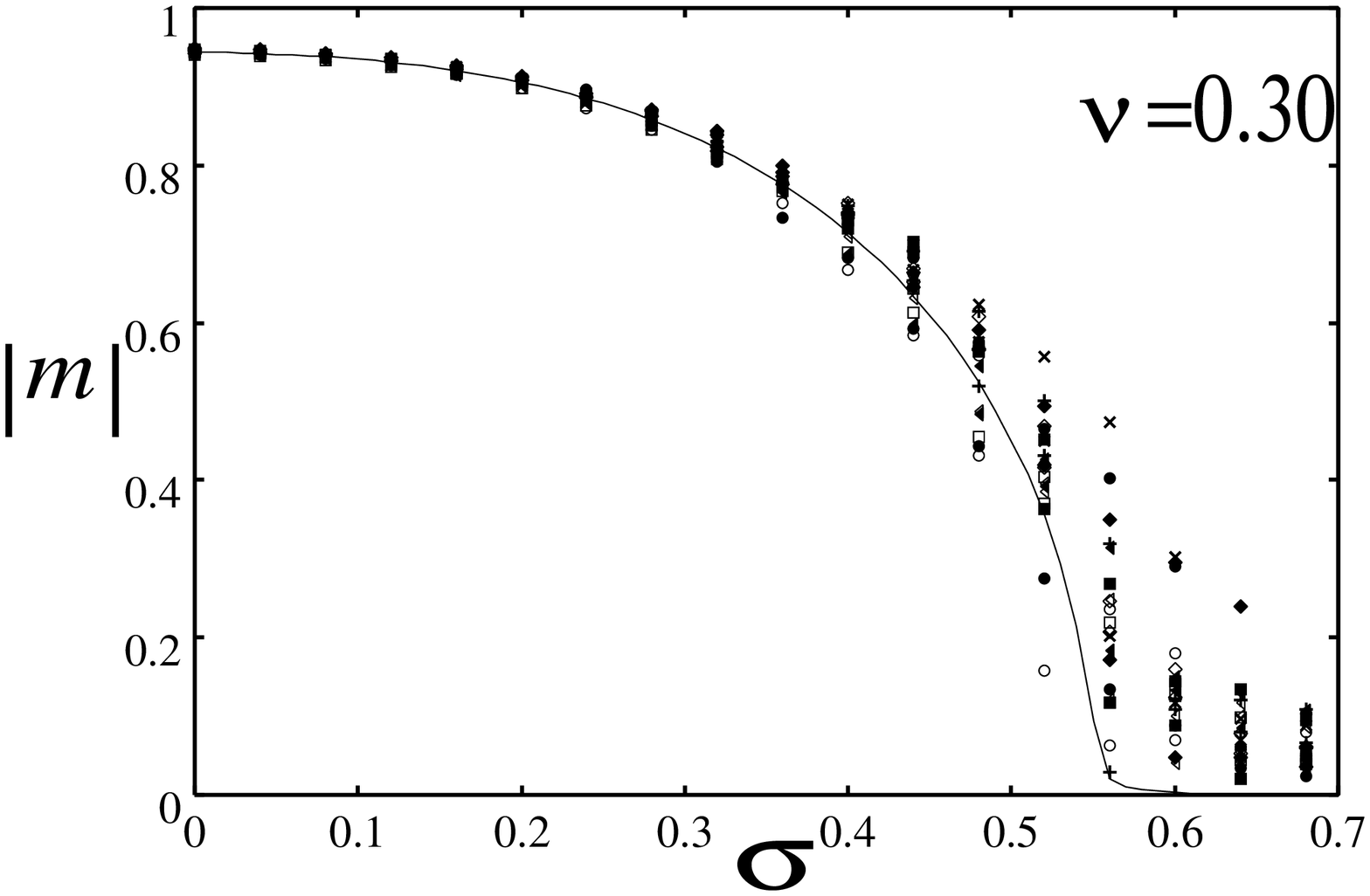}\\
\includegraphics[width=4cm]{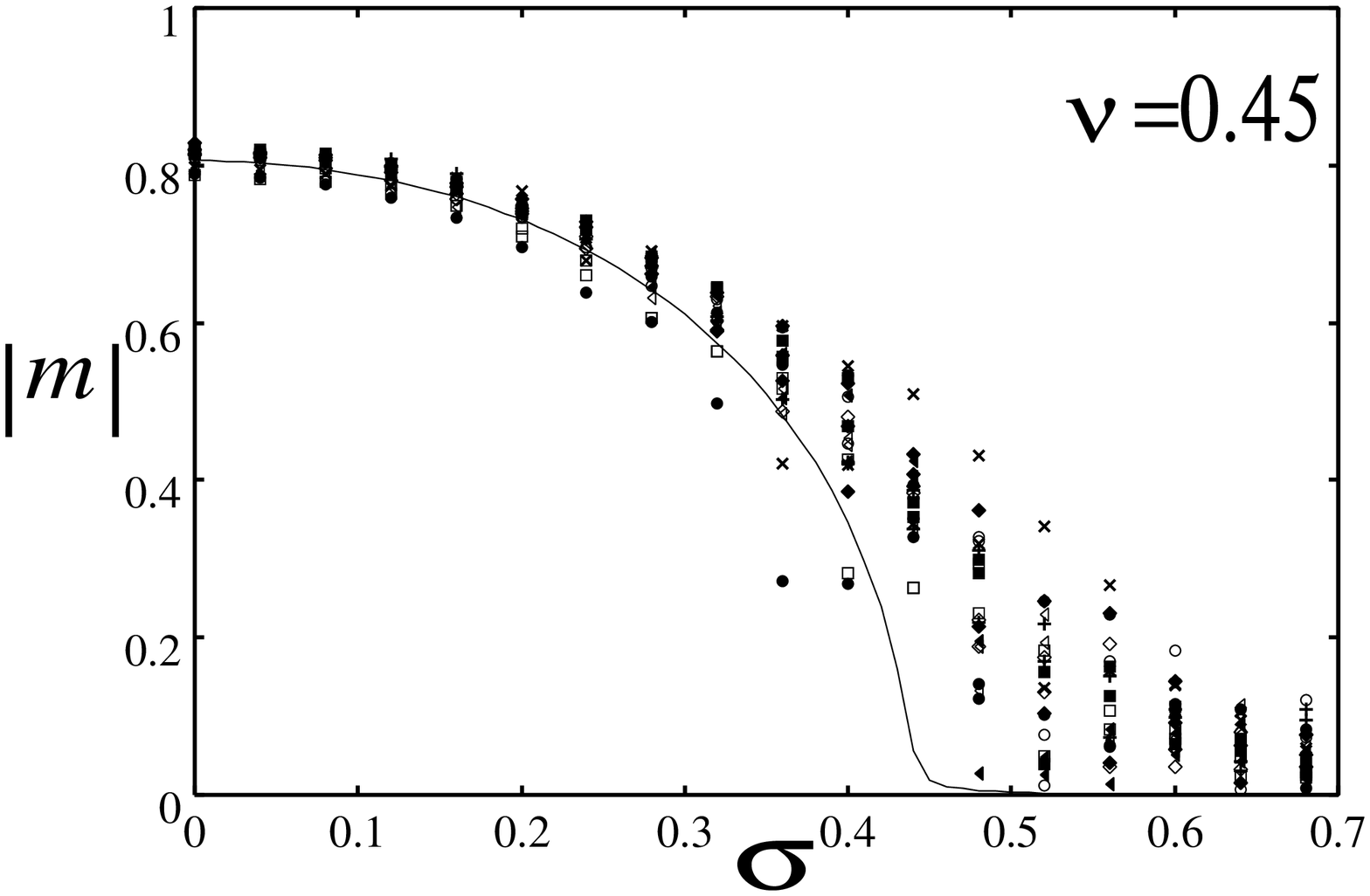}
\includegraphics[width=4cm]{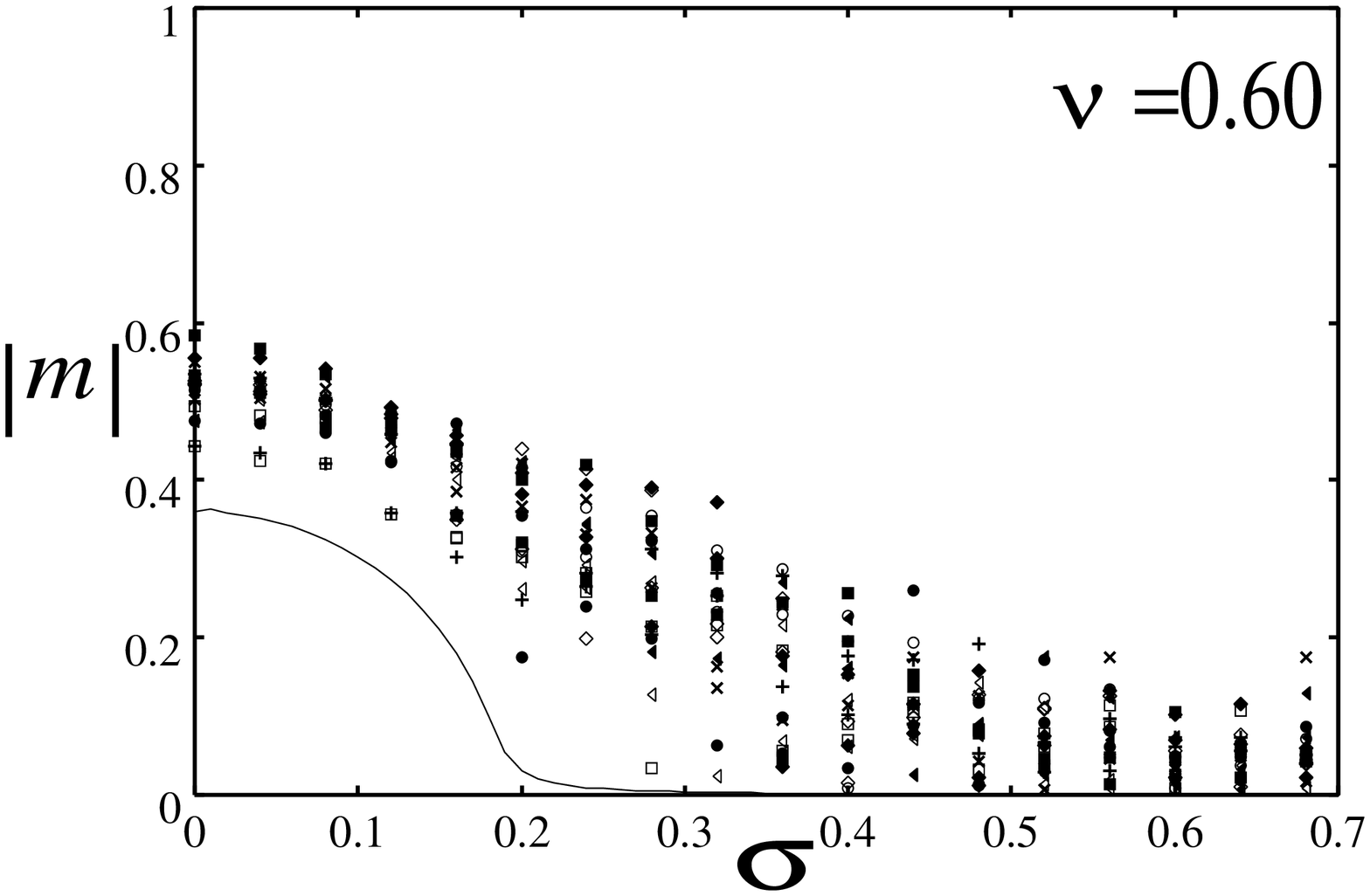}
\caption{\mbox{$|m^1|$} as a function of \mbox{$\sigma$} (solid curves were 
theoretically obtained; plots were obtained by numerical simulation). \mbox{$\beta_0= 
0$}, \mbox{$N=2000$}.}
\label{sim}
\end{figure}

\begin{figure}
(a)\includegraphics[width=3.5cm]{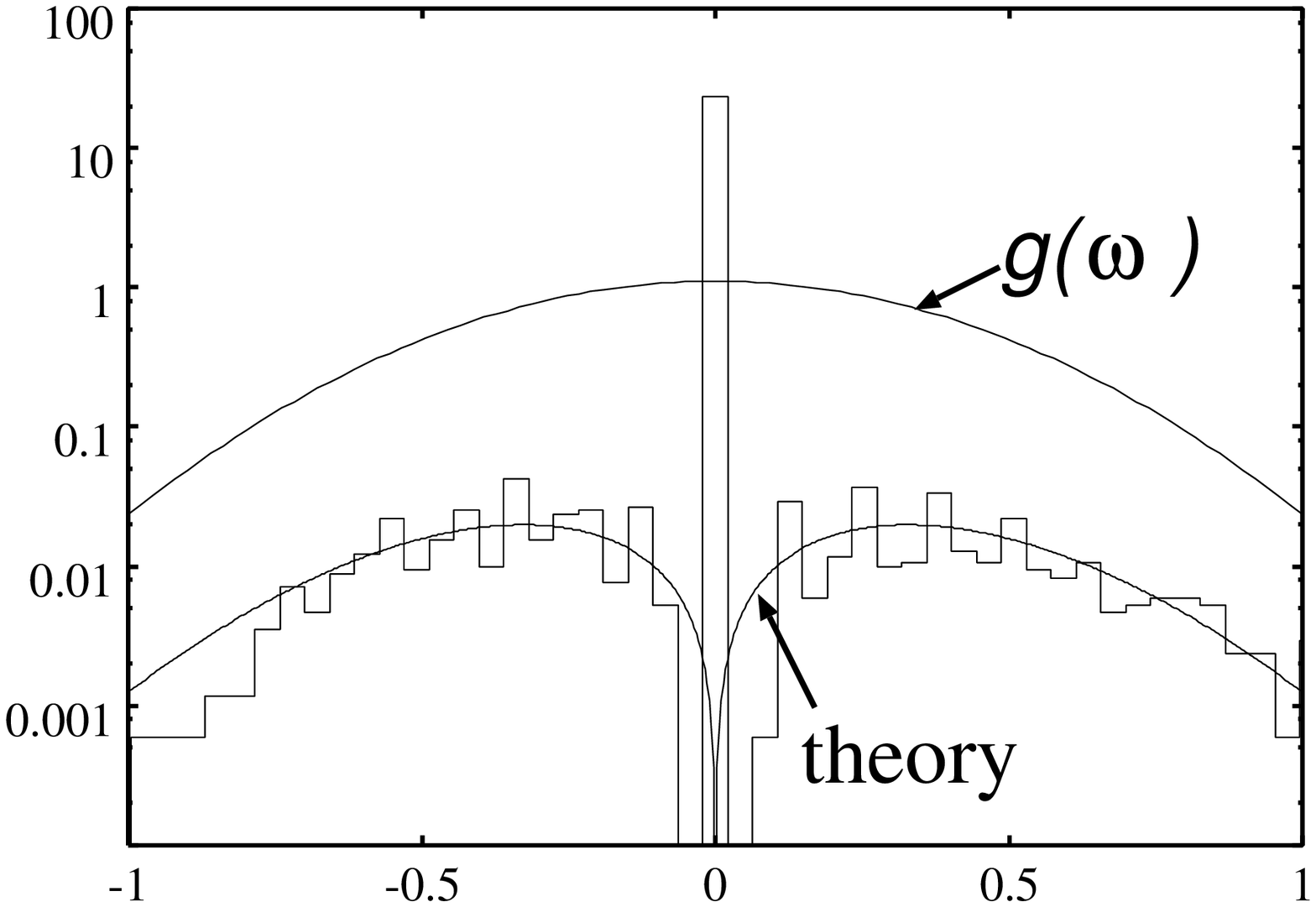}
(b)\includegraphics[width=3.5cm]{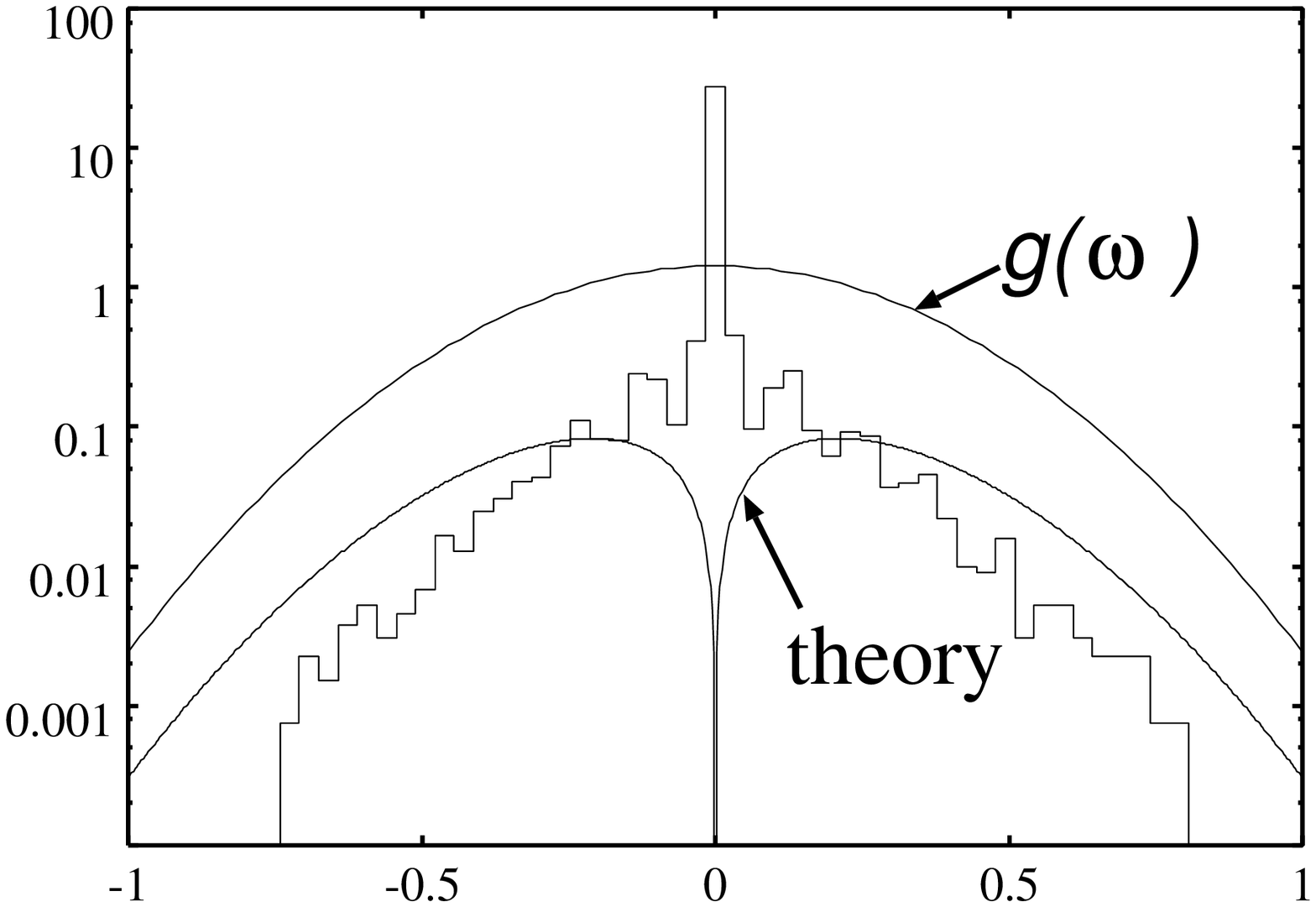}
\caption{Distribution of resultant frequencies
\mbox{$\overline{\omega}_i$} in ferromagnetic states. 
(a) \mbox{$\nu = 0.15$}, \mbox{$\sigma = 0.36$}, \mbox{$\beta_0 = 0$}. 
(b) \mbox{$\nu = 0.45$}, \mbox{$\sigma = 0.28$}, \mbox{$\beta_0 = 0$}.}
\label{dist}
\end{figure}

In the numerical simulations described here, we used a random symmetric
coupling system instead of the symmetric diluted system. We randomly
chose those couplings using two probability functions: $J_{ij}
\cos(\beta_{ij}) \sim {\cal N}(1/N, \nu^2/N)$ and $J_{ij}
\sin(\beta_{ij}) \sim {\cal N}(1/N, \nu^2/N)$,  where we restrict
mutual couplings to symmetric ones, $J_{ij} \exp(i
\beta_{ij})=J_{ji}\exp(-i \beta_{ji})$ 
and set $\beta_0=0$.  Figure \ref{phasediagram2} shows a phase diagram in the
($|m^1|$, $\sigma$, $\nu$) space, which was obtained by numerically
solving the order parameter equation (\ref{eq:og}). A cross-section of
this curved surface at $\nu=0$ is equal to a result of the 
SK theory \cite{kuramoto}. Figures \ref{sim}(a), (b), (c), and (d)
display $|m^1|$ as a function of $\sigma$ for various values of $\nu$;
the solid curves were obtained theoretically, and the data points show
results obtained by numerical simulation. Figures \ref{dist}(a) and (b)
show the distributions of the resultant frequencies
$\overline{\omega}_i$ in the ferromagnetic state. As Figs. \ref{sim} and
\ref{dist} reveal, when $|m^1|$ was small, the theoretical curves did
not fit the simulation results very well. We surmise that the gap
between the simulation results and theoretical results might have been
caused by the ergodicity breaking with the ultrametric structure
of the glass states; this breaking is related to replica-symmetry
breaking \cite{Mackenzie,parisi,mezard2}.  Unfortunately, our theory
does not capture the ultrametric structure of the glass states; it
focuses only on one of the pure states in the phase space. In the next
section, we will explain the glass phase in detail.

\subsection{Properties of spurious memory state}

Here, we examine the properties of spurious memory states. In the
following analyses, we set $c = 0.5$, $\alpha = 0.025$, $\sigma = 0.16$,
and $\beta_0 = 0$.

\begin{figure}
(a)\includegraphics[width=3.5cm]{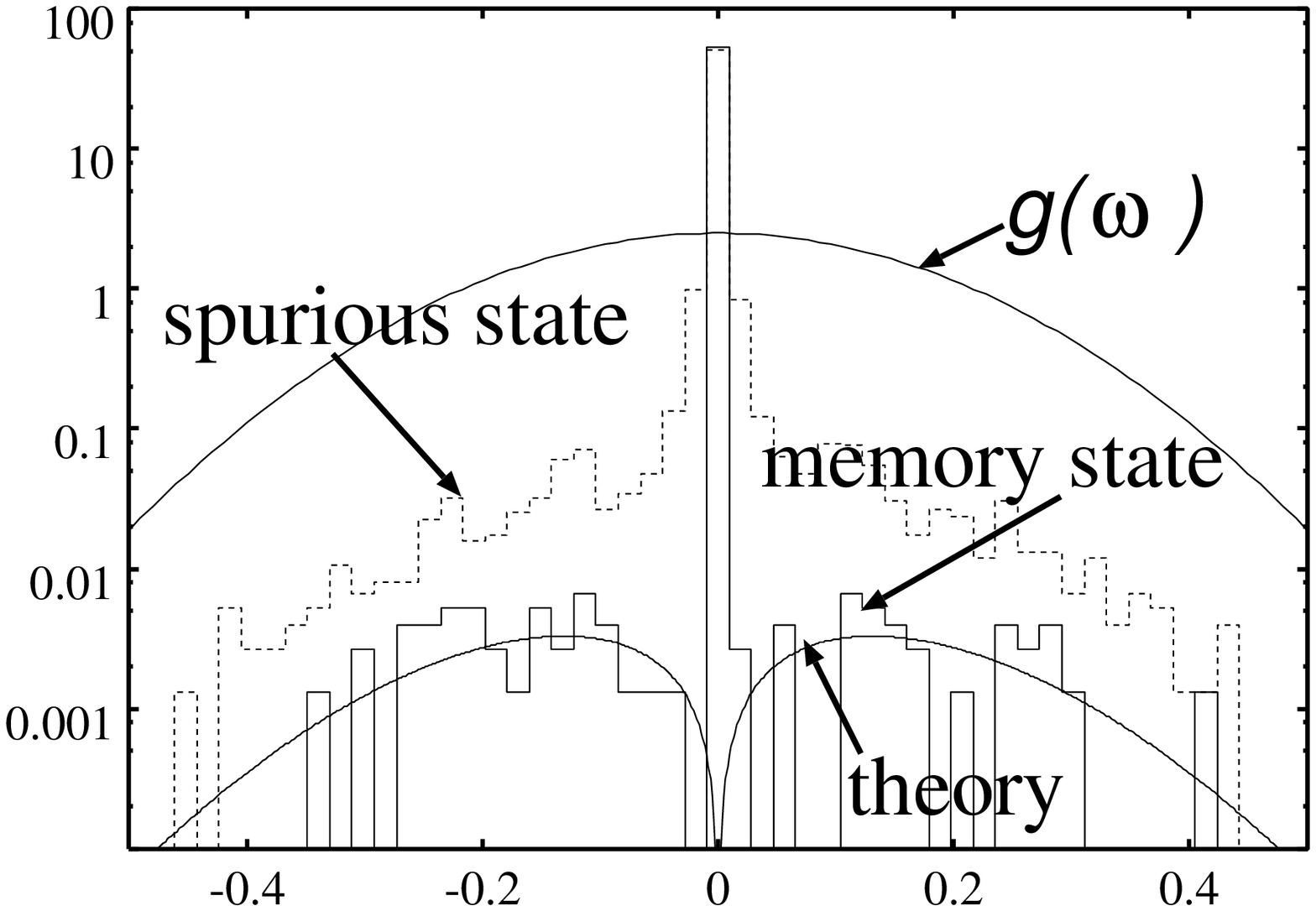}
(b)\includegraphics[width=3.5cm]{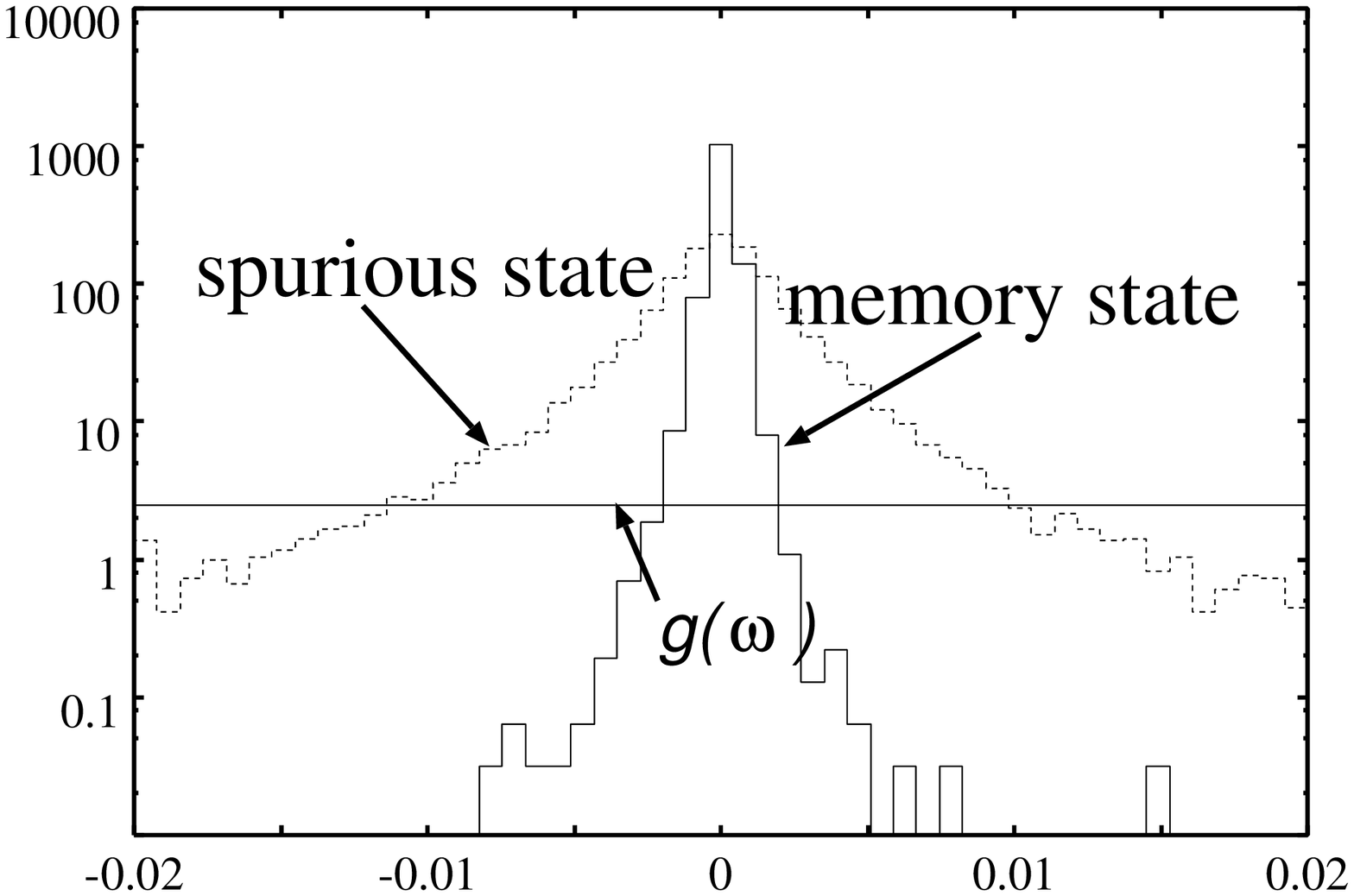}\\
(a')\includegraphics[width=3.5cm]{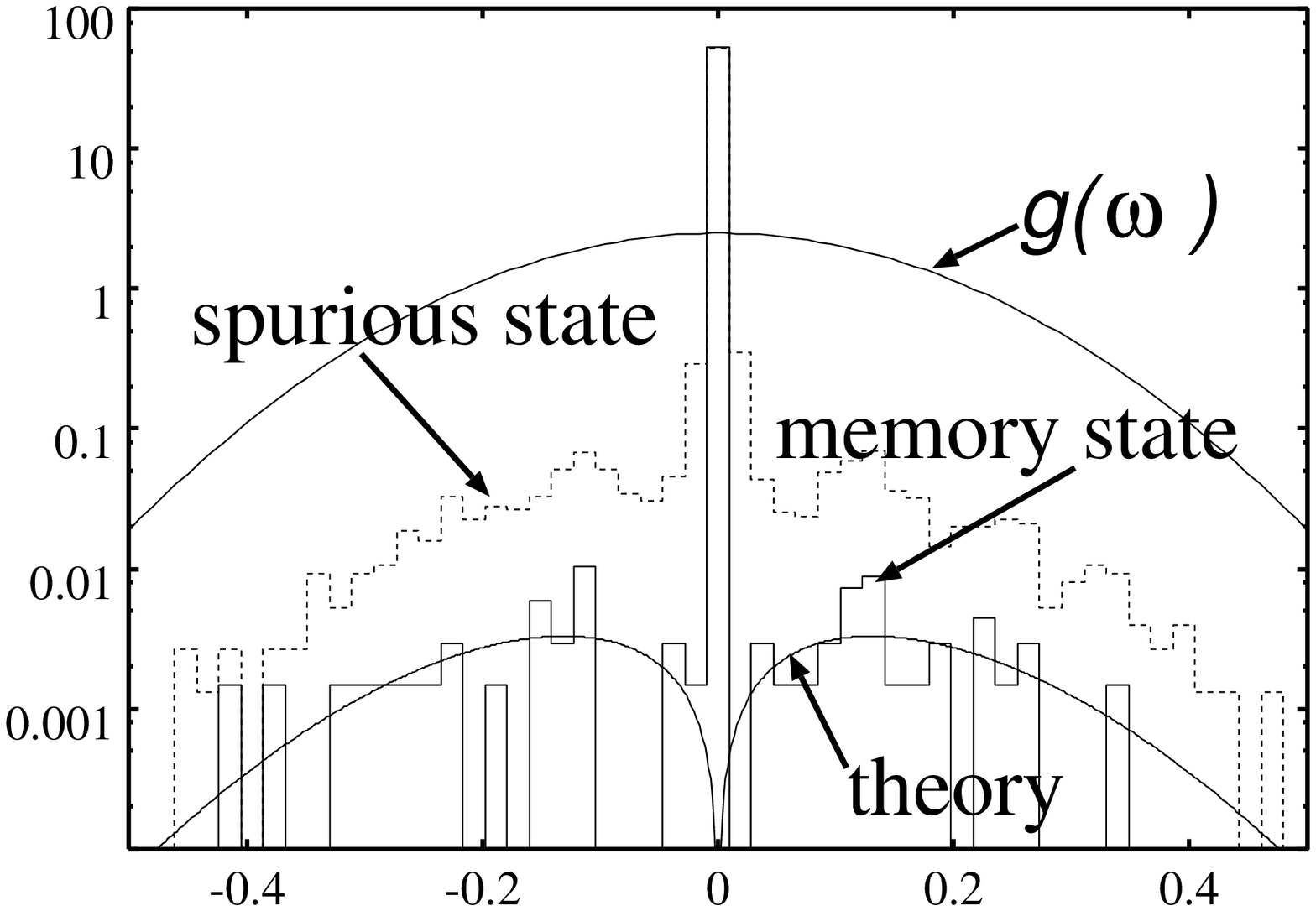}
(b')\includegraphics[width=3.5cm]{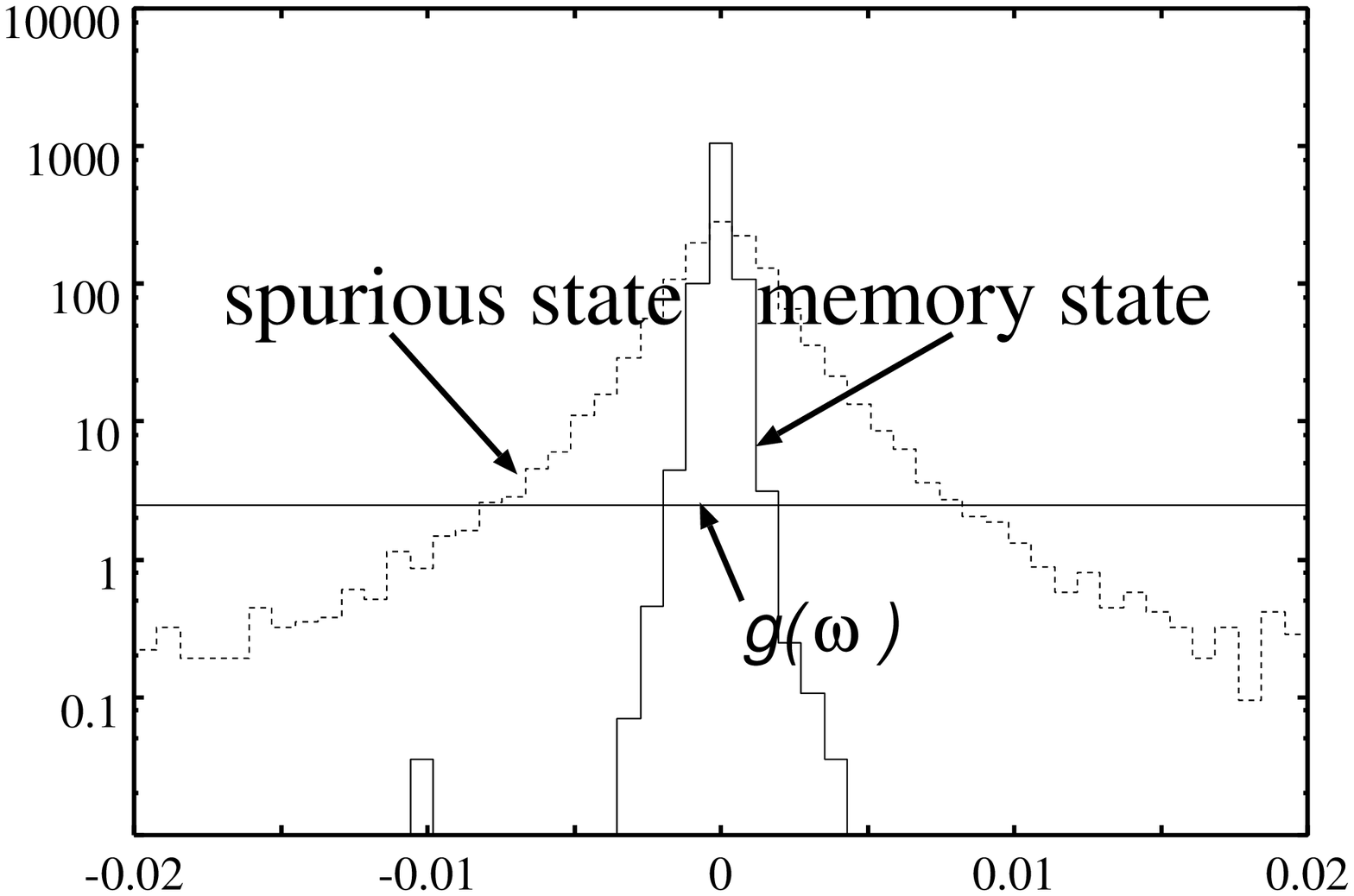} 
\caption{Distribution of
resultant frequencies $\overline{\omega}_i$ in memory and spurious
memory states.\mbox{$c = 0.5$}, \mbox{$\alpha = 0.025$}, \mbox{$\sigma =
0.16$}, and \mbox{$\beta_0 = 0$}; (b) and (b') show in detail the
structure of the sharp peaks at the center $\overline{\omega}=0$ of
\ref{distribution_b} in (a) and (a'), respectively.  (a) and (b)
asymmetry dilution; (a') and (b') symmetry dilution.}
\label{distribution_b}
\end{figure}

Figures \ref{distribution_b} (a), (a'), (b), and (b') show histograms of
resultant frequencies $\overline{\omega}_i$ in memory
states and superimpose histograms of $\overline{\omega}_i$ in spurious
memory states. Figures \ref{distribution_b} (b) and (b') show in detail
the structures of the sharp peaks at the center $\overline{\omega}=0$ in
Figs. \ref{distribution_b} (a) and (a'), respectively.  Figures
\ref{distribution_b} (a) and (b) correspond to the cases of asymmetric
diluted systems, and Figs. \ref{distribution_b} (a') and (b') correspond
to the cases of symmetric diluted systems.

As shown in Figs. \ref{distribution_b} (a) and (a'), in  
memory retrieval states, a single delta-peak exists at
$\overline{\omega} = 0$, which corresponds to a large cluster of
synchronous oscillators, and asynchronous oscillators are symmetrically
distributed around the delta-peak.

On the other hand, as shown in Figs. \ref{distribution_b} (a) and (a'),
in spurious memory states, a sharp peak exists at $\overline{\omega} =
0$. As Figs. \ref{distribution_b} (b) and (b') reveal, the peak of the
spurious memory states at $\overline{\omega} = 0$ is gentler than that
of the memory retrieval states. These gentle peak
indicates that the entrainment in the glass phase is weaker than that in
the ferromagnetic phase. This phenomenon corresponds to the so-called
{\it quasi-entrainment} observed in the glass oscillator system
\cite{daido}. As shown in Figs. \ref{distribution_b} (a) and (a'), the
 degree of asynchronous oscillators in the spurious memory states is
larger than that in the memory states. As the results
obtained from analyses of the system in the memory and
spurious memory states reveal, we can determine if the recall process is
successful or not by using information about the degrees of the
synchronous oscillators. Note that it is difficult for attractor-type
networks used to solve optimization problems to detect being trapped in
a meta-stable state during the relaxation process. This new finding
indicates that a class of non-equilibrium systems can potentially be
used to address the detection of meta-stable states.

\begin{figure}
(a)\includegraphics[width=6cm]{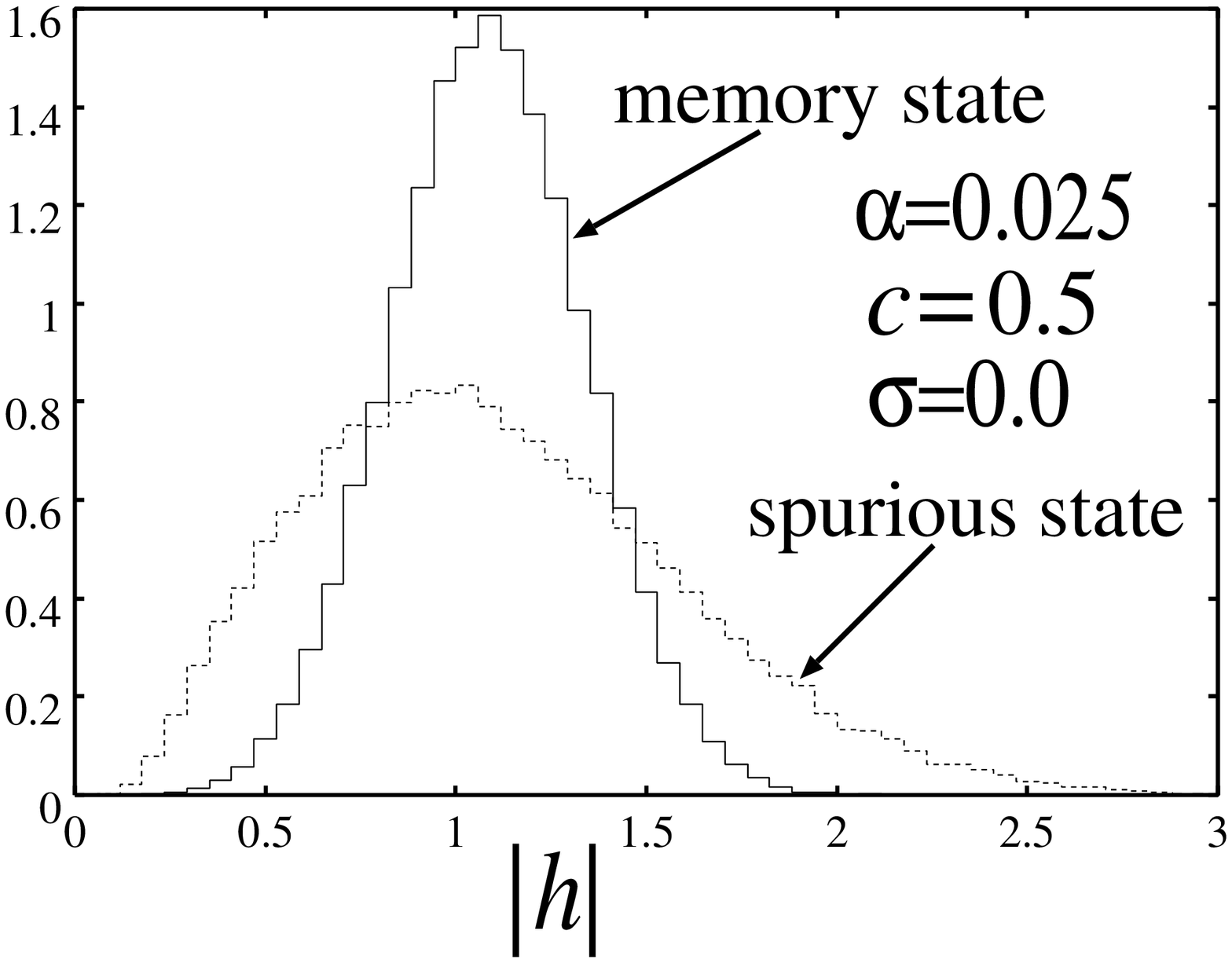}
(b)\includegraphics[width=6cm]{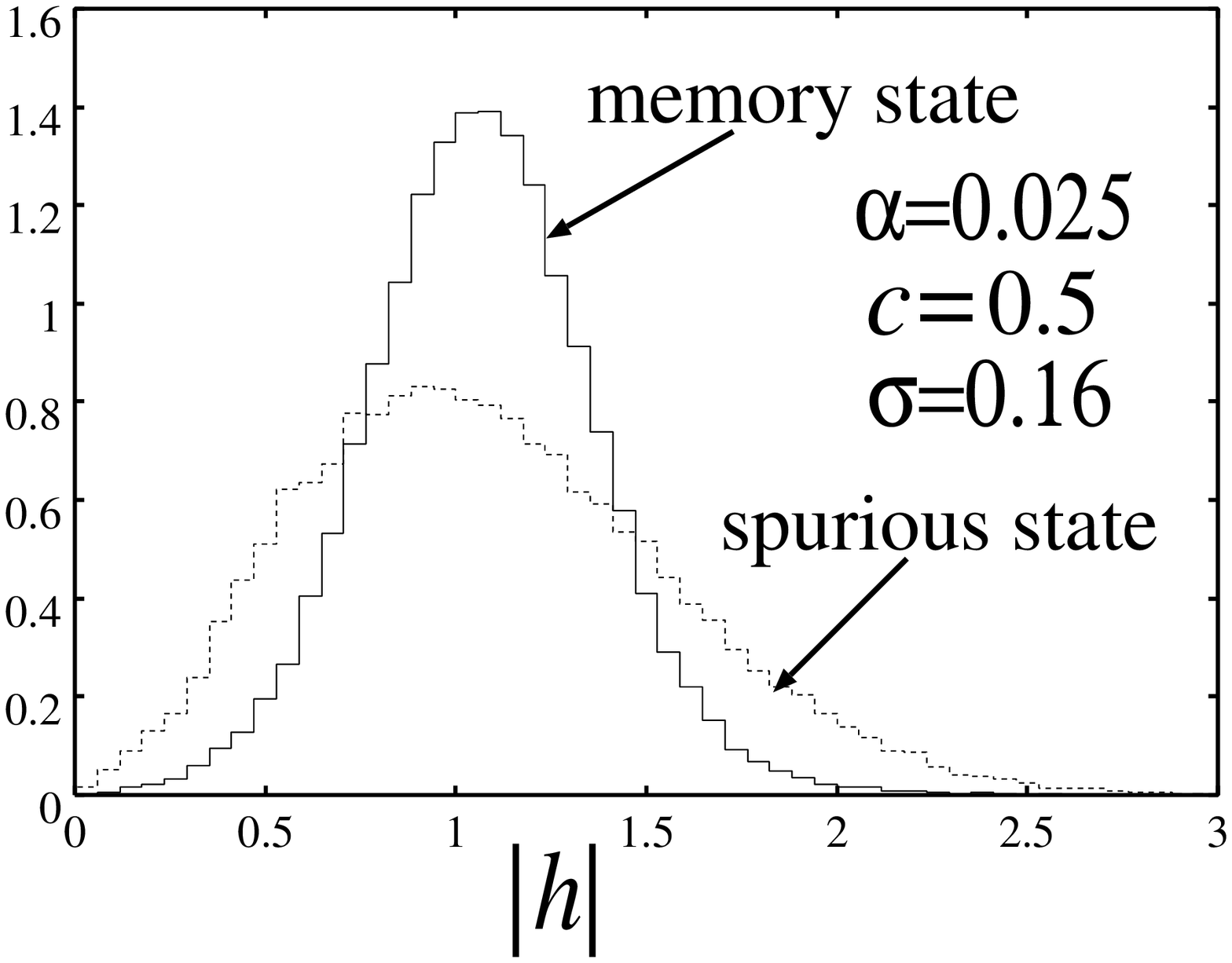}\\
(c)\includegraphics[width=6cm]{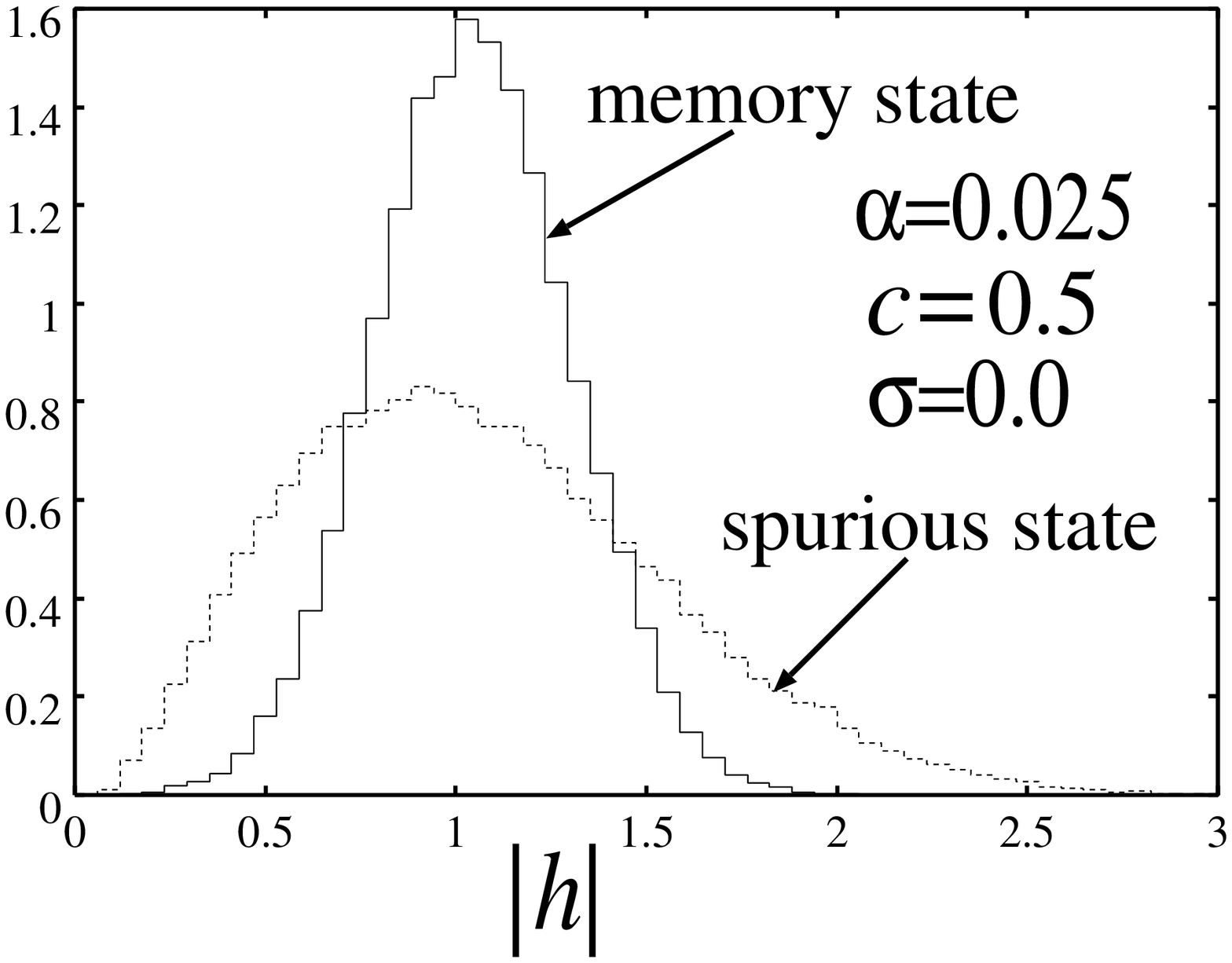}
(d)\includegraphics[width=6cm]{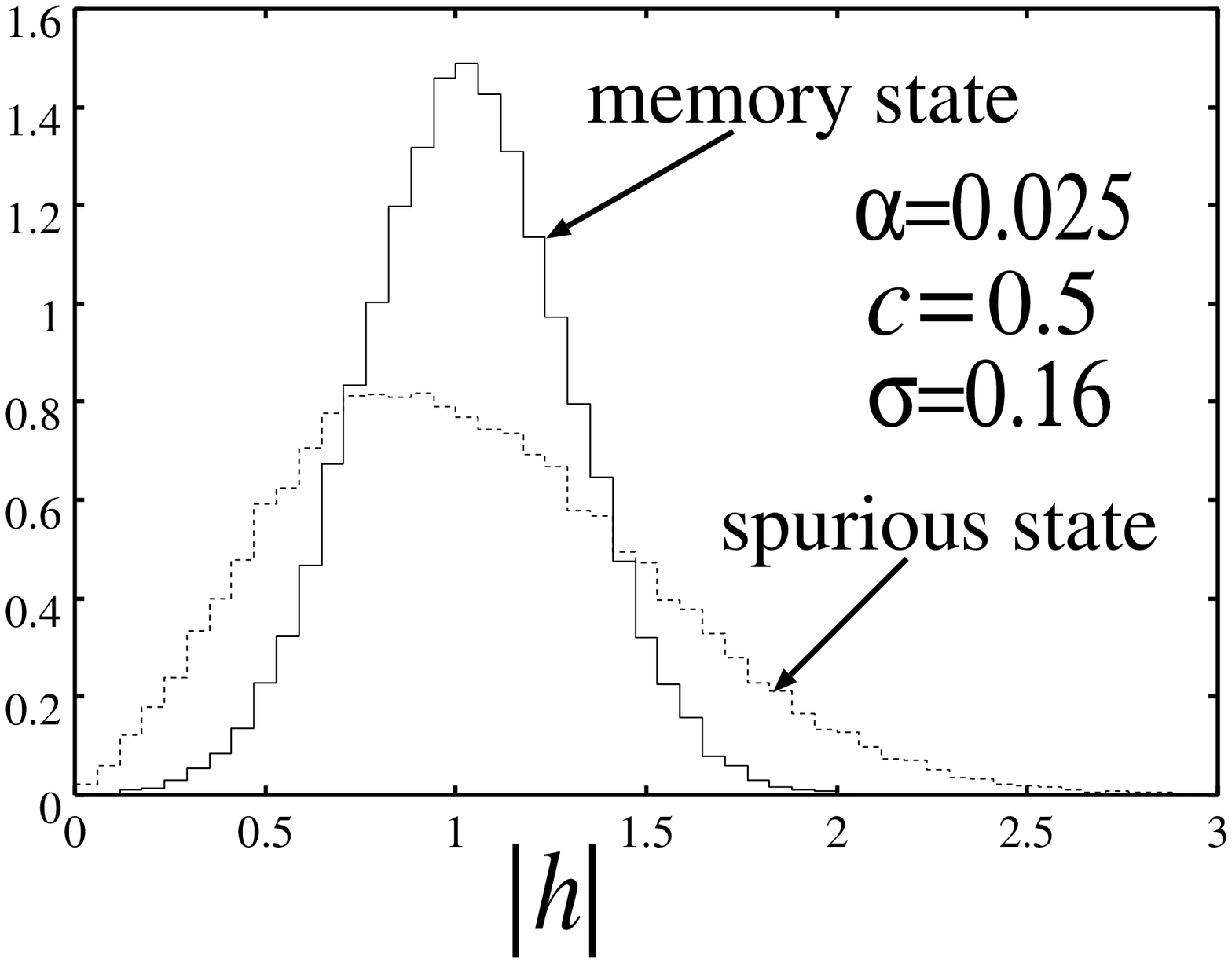} 
\caption{Distribution of local field in memory and spurious memory
states. \mbox{$\beta_0 = 0$}. (a) and (b) show results of numerical
simulation for systems with symmetry dilution; (c) and (d) show those of
systems with asymmetry dilution.} 
\label{dist2}
\end{figure}

Figures \ref{dist2}(a), (b), (c), and (d) show histograms of the
absolute value of local field $|h|$ in the memory and
spurious memory states. Since the present system possesses rotational
symmetry with respect to the phase $\phi_i$, we can safely define the
condensed pattern as $\xi^1_i=1$; i.e., the gauge transformation can be
performed on variables of the condensed pattern. After the gauge
transformation, if the system is in the memory states, the
histogram of local field $h$ is given as a two-dimensional isotropic
Gaussian at $h=m^1$ in the complex plane. However, the histogram of $h$
in the spurious memory state takes a "volcanic" form around $h=0$ in the
complex plane.  Such a histogram form for the local field was also
observed in the glass oscillator system \cite{daido}. It is well-known
that in equilibrium systems, the spin glass states have an ultrametric
tree structure \cite{Mackenzie,mezard2}. This structure of the spin
glass states can be expressed using the replica symmetric breaking
scheme in replica theory, which is based on a multi-cascade Gaussian
process for generating the local field \cite{parisi}. We surmise that
even in the non-equilibrium systems proposed here, a multi-cascade
Gaussian process in the glass states results in a non-Gaussian
distribution of the local field, as shown in Figs. \ref{dist2}(a), (b),
(c), and (d).

As mentioned above, our theory does not capture the ultrametric
structure of the glass states; it focuses only on one of the pure states
in the phase space. This is because SCSNA is based on the Gaussian
ansatz for the local field, which is deeply related to replica symmetric
approximation in replica theory \cite{fukai3}. Even in the spurious
memory state, SCSNA and replica theory under replica symmetric
approximation give the probability distribution of the local field as a
single Gaussian at $h=0$ in the complex plane. However, no one has been
able to explain the ultrametric structure of the glass states 
in the SCSNA framework. Therefore, to properly analyze the systems in
spurious states (glass states), we need to extend the present theory to
a more general theory, one that treats an ultrametric structure.

\section{Conclusion}

We have proposed a curved isochron clock (CIC) based on the radial
isochron clock that provides a clean example of the acceleration
(deceleration) effect. By analyzing two-body system of coupled CICs, we
showed that an unbalanced mutual interaction caused by curved isochron
sets is the minimum mechanism needed for
generating the acceleration (deceleration) effect
in coupled oscillator systems. From this we determined that the
Sakaguchi and Kuramoto (SK) model, which is a mean field model of
coupled oscillators without frustration, has such a mechanism.  To study
frustrated coupled oscillator systems, we extend the SK model to two
oscillator associative memory models, one with symmetric and one with
asymmetric dilution of coupling, which also have the minimum mechanism
for the acceleration (deceleration) effect.  We
theoretically showed that the {\it Onsager reaction term} (ORT), which
is unique to frustrated systems, plays an important role in the
acceleration (deceleration) effect. Comparing the two models,
we extracted the effect of the ORT only to the rotation speed of the
oscillators.

The acceleration (deceleration) effect caused by the ORT is peculiar to
non-equilibrium systems, since this effect only occurs when $\beta_0
\neq 0$. There has been fundamental disagreement regarding the existence
of the ORT in a typical system corresponding to our model with
$\beta_0=0$ and a symmetric $g(\omega)$\cite{aonishi2,yoshioka}. From
the results of this work we conclude that even if $\beta_0=0$ and
$g(\omega)$ is symmetric, the ORT exists in the bare local field given
by Eq. (\ref{eq:lf1}). In this case, the effect of the ORT is not
detectable because it cancels out of Eq. (\ref{eq:hei2}).

As the results illustrated in Figs \ref{overlap3}(a) and (b) reveal, the
critical memory capacity of the asymmetric diluted systems obtained from
numerical simulation is slightly smaller than that of the symmetric
ones, because asymmetric dilution breaks the detailed balance of the
system, which weakens the stability of the memory states. There is yet
no theory to rigorously treat a system with asymmetric interaction. Most
theoretical studies of asymmetric systems are based on the naive
assumption that there are such steady states as equilibrium states of
symmetric systems \cite{okada}.

In the field of neuroscience, a growing number of researchers are
becoming interested in the synchrony of oscillatory neural activities
because physiological evidence of their existence has been obtained in
the visual cortex of a cat \cite{Eckhorn1988,Gray1989}. Much
experimental and theoretical research has been done on the functional
role of synchronization. One of the more interesting hypotheses is
called {\it synchronized population coding}, which was proposed by
Phillips and Singer \cite{Phillips}. However, its validity is highly
controversial. In this paper, we numerically showed the possibility of
determining if the recall process was successful or not by using
information about the synchrony/asynchrony. If we consider information
processing in brain systems, the solvable {\it toy} model presented in
this paper may be a good candidate for showing the validity of a
synchronized population coding in the brain. According to anatomical and
physiological data, the olfactory system can be considered as an
associative memory with oscillatory behavior. There is one particularly
interesting finding related to our studies. Freeman and Sharda
demonstrated chaotic behavior of olfactory neural systems in response to
unknown odors \cite{freeman}. We suspect that this chaotic behavior is
related to the asynchronous behavior of our systems in spurious memory
states. Thus, the present analysis should strongly affect the debate on
the functional role of synchrony.

\section*{Appendix: Derivation of order parameter equations}

Assuming a pure effective local field, $\tilde{h}_i$, in
Eq. (\ref{eq:as2}), we performed renormalization of the local field
expressed in Eq. (\ref{eq:hei2}). The quantity 
$\tilde{\Omega}$ in Eq. (\ref{eq:hei2}) is the renormalized version of
$\Omega$, from which the ORT has been removed. Thus, $\tilde{h}_i$ is
independent of any macroscopic configuration of unit $i$. In this way,
by performing renormalization of the local field, we can reduce large
population systems to one-body problems. Under the assumption formalized
in Eq. (\ref{eq:as2}), the distribution of $s_i$, which is denoted as
$n(\phi_i; \tilde{h}_i, \tilde{\Omega})$, can be formally derived using
SK theory.

In the framework of the SK theory, we can split the distribution of
$s_i$ into a synchronized part and a desynchronized part as $n(\phi_i;
\tilde{h}_i, \tilde{\Omega})=n_s(\phi_i; \tilde{h}_i,
\tilde{\Omega})+n_{\rm ds}(\phi_i; \tilde{h}_i,
\tilde{\Omega})$. $n_s(\phi_i; \tilde{h}_i, \tilde{\Omega})$ represents
the distribution of synchronous oscillators and $n_{\rm ds}(\phi_i;
\tilde{h}_i,\tilde{\Omega})$ denotes the distribution of asynchronous
ones.  First, we derive $n_s(\phi_i; \tilde{h}_i, \tilde{\Omega})$ from
Eq. (\ref{eq:hei2}) ($\frac{d \phi_i}{d \tau}=0$) as follows,
\begin{widetext}
\begin{eqnarray}
n_s(\phi_i; \tilde{h}_i, \tilde{\Omega}) &=& 
 \int d \omega_i g(\omega_i)\delta\left( \omega_i - \tilde{\Omega} - \sin(\phi_i) 
\tilde{h}^R_i +\cos(\phi_i) \tilde{h}^I_i\right)\nonumber \\
&=&g \left(\tilde{\Omega} + \tilde{h}^R_i \sin(\phi_i) - \tilde{h}^I_i 
\cos(\phi_i)\right) 
\left(\tilde{h}^R_i \cos(\phi_i)+ \tilde{h}^I_i \sin(\phi_i) \right),\\
& &-\frac{\pi}{2} + \tan^{-1}\frac{\tilde{h}^I_i}{\tilde{h}^R_i}  \leq \phi_i \leq 
\frac{\pi}{2} + \tan^{-1}\frac{\tilde{h}^I_i}{\tilde{h}^R_i} \nonumber.
\end{eqnarray}
Next, we consider $n_{ds}(\phi_i; \tilde{h}_i, \tilde{\Omega},\omega_i)$ 
which represents a conditional probability distribution of $n_{\rm ds}(\phi_i; 
\tilde{h}_i, \tilde{\Omega})$.
$n_{\rm ds}(\phi_i; \tilde{h}_i, \tilde{\Omega}, \omega_i)$ is governed by the 
following Liouville equation,
\begin{eqnarray}
\frac{\partial}{\partial \tau}n_{ds}(\phi_i; \tilde{h}_i, \tilde{\Omega},\omega_i)=
-\frac{\partial}{\partial \phi_i}
\left((\omega_i-\tilde{\Omega}-\sin(\phi_i)\tilde{h}^R_i +
\cos(\phi_i)\tilde{h}^I_i)n_{ds}(\phi_i; \tilde{h}_i, \tilde{\Omega}, \omega_i)
\right).
\end{eqnarray}
In the limit that $\tau \rightarrow
\infty$, the stationary distribution becomes 
\begin{eqnarray}
n_{ds}(\phi_i; \tilde{h}_i, \tilde{\Omega},\omega_i) &=& 
C(\omega_i-\tilde{\Omega}-\sin(\phi_i)\tilde{h}^R_i +
\cos(\phi_i)\tilde{h}^I_i)^{-1} \nonumber \\
C &=& 1/\int_0^{2\pi} d \phi_i 
(\omega_i-\tilde{\Omega}-\sin(\phi_i)\tilde{h}^R_i 
+\cos(\phi_i)\tilde{h}^I_i)^{-1} \nonumber \\
 &=& \frac{\omega_i - \tilde{\Omega}}{2\pi} \sqrt{1 - 
\frac{|\tilde{h}_i|^2}{(\omega_i-\tilde{\Omega})^2}}.
\end{eqnarray}
Then, $n_{\rm ds}(\phi_i; \tilde{h}_i, \tilde{\Omega})$ is expressed as
\begin{eqnarray}
n_{ds}(\phi_i; \tilde{h}_i, \tilde{\Omega}) = \frac{1}{2\pi}
\int_{|\omega_i-\tilde{\Omega}| > |\tilde{h}_i|} d\omega_i g(\omega_i)
\frac{(\omega_i-\tilde{\Omega})\sqrt{1 - 
\frac{|\tilde{h}_i|^2}{(\omega_i-\tilde{\Omega})^2}}}{\omega_i-\tilde{\Omega}
-\sin(\phi)\tilde{h}^R_i + \cos(\phi_i)\tilde{h}^I_i}.
\end{eqnarray}
Averaging $s_i$ over $\omega_i$, that is, $\left<s_i\right>_{\omega_i} = \int
d\phi_i n(\phi_i; \tilde{h}_i, \tilde{\Omega})\exp(i \phi_i)$, we obtain the 
following equation,
\begin{eqnarray}
\left< s_i(\tilde{h}_i) \right>_{\omega_i} &=& \tilde{h}_i
\int_{-\pi/2}^{\pi/2} d \phi_i g\left(\tilde{\Omega} + |\tilde{h}_i|\sin\phi_i
\right) \cos\phi_i \exp(i \phi_i) \nonumber \\
&+& i \frac{\tilde{h}_i}{|\tilde{h}_i|}
 \int_{|\omega_i-\tilde{\Omega}| > |\tilde{h}_i|} d\omega_i
g(\omega_i)(\omega_i-\tilde{\Omega})\left(1-\sqrt{1-\frac{|\tilde{h}_i|^2}{(\omega_i-\tilde{\Omega})^2}}\right)\nonumber 
\\
&=& \tilde{h}_i \int_{-\pi/2}^{\pi/2} d \phi_i g\left(\tilde{\Omega} +
|\tilde{h}_i|\sin\phi_i \right)\cos\phi_i\exp(i \phi_i)\nonumber \\ 
&+& i  \tilde{h}_i \int_{0}^{\pi/2} d \phi_i
\frac{\cos\phi_i (1-\cos\phi_i)}{\sin^3\phi_i} \left\{ 
g\left(\tilde{\Omega}+\frac{|\tilde{h}_i|}{\sin\phi_i}\right)-g\left(\tilde{\Omega
}-\frac{|\tilde{h}_i|}{\sin\phi_i}\right)
\right\}. \label{eq.X}
\end{eqnarray}
\end{widetext}
Note that if $g(\tilde{\Omega}+x)=g(\tilde{\Omega}-x)$, we can neglect the effect of 
asynchronous oscillators. When $g(x)=\delta(x)$ and $\beta_0 = 0$, we obtain 
\begin{eqnarray}
\left<s_i(\tilde{h}_i)\right>_\omega &=& \frac{\tilde{h}_i}{|\tilde{h}_i|}. 
\end{eqnarray}

Next, we estimate $\tilde{h}_i$ in the framework of SCSNA. In this
analysis, we focus on the memory retrieval states, in which the
configuration has appreciable overlap with the condensed pattern
$\mbox{\boldmath $\xi$}^1$ ($m^1 \sim O(1)$) and has little overlap with
the uncondensed patterns $\mbox{\boldmath $\xi$}^\mu$ for $\mu>1$
($m^\mu \sim O(1/\sqrt{N})$).  Under this assumption, we estimate the
contribution of the uncondensed patterns using SCSNA \cite{fukai3} and
determine $\tilde{h}_i$ in a self-consistent manner.  In the first step
of SCSNA, we split local field $h_i$ into a signal part (the first
term), a cross-talk noise part (the second term), and a coupling noise
part (the third term):
\begin{eqnarray}
h_i e^{-i\beta_0} = \xi_i^{1} m^1 + \sum_{\mu>1} \xi_i^{\mu} m^\mu 
+ \sum_{j(\neq i)}^{N} \delta n_{ij} s_j - \alpha s_i. \label{eq:SN}
\end{eqnarray}
In the next step, we split the cross-talk noise (the second term) and
the coupling noise (the third term), respectively, into the Gaussian
random variable and the ORT. Equation (\ref{eq:hei1}) implies that $s_i$
is a function of local field $h_i$, natural frequency $\omega_i-\Omega$,
and time $\tau$; that is, 
\begin{equation}
 s_i = X(h_i, \omega_i-\Omega, \tau).
\end{equation}
Note that $s_i$ is not a function of renormalized $\tilde{h}_i$ and
$\tilde{\Omega}$; instead, it is a function of the bare $h_i$ and
$\Omega$ in Eq. (\ref{eq:hei1}).  We can properly evaluate the ORT with
this careful treatment. Here, we assume that the microscopic memory
effect can be neglected in the $\tau \rightarrow \infty$ limit. In
general, $X(h_i, \omega_i-\Omega, \tau)$ is not regular. As such, a
variation in $X$ due to a small perturbation in local field $h$ denoted
$dh$ is satisfied
\begin{equation}
 d X = u(h, \omega-\Omega, \tau) dh + v(h, \omega-\Omega, \tau) d {\overline h}.
\end{equation}
Therefore, $m^\mu \sim O(1/\sqrt{N})$, $\mu \geq 2$ is expressed as 
\begin{eqnarray}
 m^\mu &=& \frac{1}{N} \sum_i  {\overline\xi}_i^\mu X(h_i, \omega_i-\Omega, 
\tau) \nonumber \\
&=& \frac{1}{N} \sum_i {\overline\xi}_i^\mu X_i^{(\mu)}
  + U_1 m^\mu e^{i \beta_0}+ V_1 {\overline m}^\mu e^{-i\beta_0}, 
\label{eq:hiki1}  \\
 U_1 &=& \frac{1}{N} \sum_i {\overline\xi}_i^\mu \xi_i^\mu u(h^{(\mu)}_i, 
\omega_i-\Omega, \tau), \\
 V_1 &=& \frac{1}{N} \sum_i {\overline\xi}_i^\mu {\overline\xi}_i^\mu 
v(h^{(\mu)}_i, \omega_i-\Omega, \tau),
\end{eqnarray}
where
\begin{eqnarray}
& & X_i^{(\mu)} = X\left( h_i^{(\mu)}, \omega_i-\Omega, \tau \right), \nonumber 
\\ 
& & h_i^{(\mu)}e^{-i\beta_0}=\sum_{\nu(\neq\mu)} \xi_i^\nu m^\nu + 
\sum_{j(\neq i)}^{N} \delta n_{ij} s_j  - \alpha s_i.
\end{eqnarray}
Note that $X_i^{(\mu)}$ is uncorrelated with $\xi_i^\mu$. 
We can neglect the complex conjugate term $V_1$, which
leads to a higher-harmonic term of the coupling function \cite{aonishi},
since $E\left[{\overline\xi}_i^\mu {\overline\xi}_i^\mu\right]=0$.

From Eq. (\ref{eq:hiki1}), we obtain
\begin{eqnarray}
 m^\mu &=& \frac{1}{N(1- e^{i \beta_0}U_1)} 
  \sum_j {\overline\xi}_j^\mu X_j^{(\mu)}. \label{eq:ucov}
\end{eqnarray}
Substituting Eq. (\ref{eq:ucov}) into the cross-talk noise of
Eq. (\ref{eq:SN}), in the limit $N \rightarrow \infty$, we can split the
cross-talk noise into the Gaussian random variable and the ORT:
\begin{eqnarray}
\sum_{\mu=2}^{\alpha N} \xi_i^{\mu} m^\mu &=& z^A_i 
  + \frac{\alpha}{1- e^{i \beta_0}U} s_i, \\
z^A_i &=& \frac{1}{N(1- e^{i \beta_0}U)} \sum_{\mu=1}^{\alpha N}
  \sum_{j(\neq i)} \xi_i^\mu {\overline\xi}_j^\mu X_j^{(\mu)}, \\
U &=& \frac{1}{N} \sum_i u(h_i, \omega_i-\Omega, \tau),
\end{eqnarray}
Next, we split the coupling noise term of Eq. (\ref{eq:SN}) into the
Gaussian random variable and the ORT.
\begin{eqnarray}
\sum_{j \neq i}^{N} \delta n_{ij} s_j &=& \sum_{j(\neq i)}^{N} \delta n_{ij}
X_j^{(i)} \nonumber \\
&+& s_i  e^{i \beta_0} U_2 +  \overline{s}_i e^{-i\beta_0} V_2, \\
U_2 &=& \frac{1}{N} \sum_{j (\neq i)}^{N} \delta n_{ij} \delta n_{ji}
u(h^{(i)}_j, \omega_j-\Omega, \tau), \\
V_2 &=& \frac{1}{N} \sum_{j (\neq i)}^{N} \delta n_{ij} \overline{\delta n_{ji}} 
v(h^{(i)}_j, \omega_j-\Omega, \tau),
\end{eqnarray}
where
\begin{eqnarray}
& & X_i^{(j)} = X\left( h_i^{(j)}, \omega_i-\Omega, \tau \right), \nonumber \\ 
& & h_i^{(j)}e^{-i\beta_0}=\sum_{\nu} \xi_i^\nu m^\nu + \sum_{k(\neq i, 
j)}^{N} \delta n_{ik} s_k - \alpha s_i,
\end{eqnarray}
Note that $X_i^{(j)}$ is uncorrelated with $\delta n_{ij}$. We can
also neglect the complex conjugate term $V_2$, which leads to a
higher-harmonic term of the coupling function \cite{aonishi}, since
$E[\delta n_{ij} \overline{\delta n_{ji}}] = 0$.  For the symmetric
diluted system, $E[\delta n_{ij} \delta n_{ji}] = \frac{2\nu^2}{N}$, so,
in the limit $N \rightarrow \infty$, we can split the coupling noise
term of Eq. (\ref{eq:SN}) into the Gaussian random variable and the ORT:
\begin{eqnarray}
 \sum_{j=1}^{N} \delta n_{ij} s_j &=& z^G_i + 2\nu^2  e^{i \beta_0} 
U s_i, \\
z^G_i &=& \sum_{j (\neq i)}^{N} \delta n_{ij} X_j^{(i)} \\
U &=& \frac{1}{N} \sum_{j}^{N} u(h_j, \omega_j-\Omega, \tau),
\end{eqnarray}
For the asymmetric diluted system, we can neglect the ORT in the coupling noise term:
\begin{eqnarray}
\sum_{j=1}^{N} \delta n_{ij} s_j &=& z^G_i, \\
z^G_i &=& \sum_{j (\neq i)}^{N} \delta n_{ij} X_j^{(i)},
\end{eqnarray}
since $E[\delta n_{ij} \delta n_{ji}] = 0$. 

From these manipulations, local field $h_i$ is given by
\begin{eqnarray}
& &h_i e^{-i\beta_0} = \xi^1_i m^1 + z^A_i + z^G_i + \Gamma s_i \\
& &\Gamma  = \frac{\alpha e^{i \beta_0} U}{1- e^{i \beta_0}U} +2\nu^2  e^{i 
\beta_0}U\ \  {\rm (symmetric)}, \\
& &\Gamma = \frac{\alpha e^{i \beta_0} U}{1- e^{i \beta_0}U}\ \ {\rm 
(asymmetric)},
\end{eqnarray}
where $z^A_i + z^G_i$ is a random variable of an isotropic 2-dimensional
Gaussian that satisfies
\begin{eqnarray}
E\left[{\rm Re}[z^A_i + z^G_i]^2\right] &=& E\left[{\rm Im}[z^A_i
+ z^G_i]^2\right] \nonumber\\
&=& \frac{\alpha }{2 |1- e^{i \beta_0}U|^2} + \nu^2, \nonumber \\
E\left[{\rm Re}[z^A_i + z^G_i] {\rm Im}[z^A_i + z^G_i] \right] &=& 0, \nonumber
\end{eqnarray}

In the final step of SCSNA, we determine a pure effective local field 
$\tilde{h}_i$ in a self-consistent manner. We present Eq. (\ref{eq:hei2}) again:
\begin{eqnarray}
- \frac{d \phi_i}{d\tau} + \omega_i - \tilde{\Omega} = 
\sin \phi_i  \tilde{h}_i^R  - \cos \phi_i
\tilde{h}_i^I. \label{eq:av1}
\end{eqnarray}
On the other hand, local field $h_i$ is given by
\begin{eqnarray}
h_i e^{-i\beta_0} &=& \xi^1_i m^1 + z^A_i + z^G_i + \left|\Gamma \right|
\exp(i \psi ) s_i,\label{eq:local} \\
\psi &=& {\rm Arg} \left(\Gamma \right).
\end{eqnarray}
Substituting Eq. (\ref{eq:local}) into Eq. (\ref{eq:hei1}), we obtain
\begin{widetext}
\begin{eqnarray}
- \frac{d \phi_i}{d\tau} + \omega_i - \Omega &=&  \sin\phi_i
\left( {\rm Re}[e^{i \beta_0}(\xi^1_i m^1 + z^A_i + z^G_i)]
+\left|\Gamma\right| \cos (\phi_i + \psi + \beta_0)
\right) \nonumber \\
& & - \cos \phi_i \left( {\rm Im}[e^{i \beta_0}(\xi^1_i m^1 + z^A_i + z^G_i)]
+ \left|\Gamma\right| \sin(\phi_i + \psi + \beta_0) \right)
\nonumber \\ 
&=& 
\sin \phi_i{\rm Re}[e^{i \beta_0}(\xi^1_i m^1 + z^A_i + z^G_i)] \nonumber \\
& & - \cos \phi_i {\rm Im}[e^{i \beta_0}(\xi^1_i m^1 + z^A_i +
z^G_i)] -\left|\Gamma\right|\sin(\psi + \beta_0) .  \label{eq:av2}
\end{eqnarray}
\end{widetext}
Comparing Eq. (\ref{eq:av1}) with Eq. (\ref{eq:av2}),
we can determine effective local field $\tilde{h}_i$: 
\begin{eqnarray}
\tilde{h}_i e^{-i\beta_0}= \xi^1_i m^1 + z^A_i + z^G_i, 
\end{eqnarray}
and we can obtain $\tilde{\Omega}$, which is the renormalized version of $\Omega$:
\begin{eqnarray}
\tilde{\Omega} = \Omega - \left|\Gamma \right|\sin(\psi+\beta_0). 
\end{eqnarray}
These final two results are consistent with the first assumption for a pure effective local 
field $\tilde{h}_i$ in Eq. (\ref{eq:as2}).

Finally, we combine the results obtained from SK theory and SCSNA. We
can safely replace $X(h_i, \omega_i-\Omega, \tau)$ of order parameters
$m^1$ and $U$ with Eq. (\ref{eq.X}) based on $\tilde{h}_i$:
\begin{eqnarray}
m^1 &=& \frac{1}{N} \sum_i  \overline{\xi}^1_i X(h_i, \omega_i-\Omega,\tau)
\nonumber \\ 
&=& \frac{1}{N} \sum_i \overline{\xi}^1_i \left<s_i(\tilde{h}_i) 
\right>_{\omega_i} \\
U &=& \frac{1}{N} \sum_i \frac{\partial X(h_i, \omega_i-\Omega,\tau)}{\partial 
h_i} \nonumber \\ 
&=& \frac{1}{N} \sum_i \frac{\partial  \left<s_i(\tilde{h}_i) 
\right>_{\omega_i}}{\partial \tilde{h}_i}, \\
\tilde{h}_i e^{-i\beta_0} &=& \xi^1_i m^1 + z^A_i + z^G_i, 
\end{eqnarray}
because $\tilde{h}_i$ is independent of any microscopic configuration of
unit $i$. In conclusion, we can obtain the order parameter equations
(\ref{eq.o1}) and (\ref{eq.o2}). In this way, our derivation process is
complete in a self-consistent manner.

\bibliography{ref.bib}

\begin{thebibliography}{30}
\expandafter\ifx\csname natexlab\endcsname\relax\def\natexlab#1{#1}\fi
\expandafter\ifx\csname bibnamefont\endcsname\relax
  \def\bibnamefont#1{#1}\fi
\expandafter\ifx\csname bibfnamefont\endcsname\relax
  \def\bibfnamefont#1{#1}\fi
\expandafter\ifx\csname citenamefont\endcsname\relax
  \def\citenamefont#1{#1}\fi
\expandafter\ifx\csname url\endcsname\relax
  \def\url#1{\texttt{#1}}\fi
\expandafter\ifx\csname urlprefix\endcsname\relax\def\urlprefix{URL }\fi
\providecommand{\bibinfo}[2]{#2}
\providecommand{\eprint}[2][]{\url{#2}}

\bibitem[{\citenamefont{Meunier}(1992)}]{meunier}
\bibinfo{author}{\bibfnamefont{C.}~\bibnamefont{Meunier}},
  \bibinfo{journal}{Biol. Cybern.} \textbf{\bibinfo{volume}{67}},
  \bibinfo{pages}{155} (\bibinfo{year}{1992}).

\bibitem[{\citenamefont{D.Hansel et~al.}(1993)\citenamefont{D.Hansel, Mato, and
  Meunier}}]{Hansel1}
\bibinfo{author}{\bibnamefont{D.Hansel}},
  \bibinfo{author}{\bibfnamefont{G.}~\bibnamefont{Mato}}, \bibnamefont{and}
  \bibinfo{author}{\bibfnamefont{C.}~\bibnamefont{Meunier}},
  \bibinfo{journal}{Europhysics Letters} \textbf{\bibinfo{volume}{23}},
  \bibinfo{pages}{367} (\bibinfo{year}{1993}).

\bibitem[{\citenamefont{Ermentrout}(1981)}]{ermentrout}
\bibinfo{author}{\bibfnamefont{G.~B.} \bibnamefont{Ermentrout}},
  \bibinfo{journal}{Journal of Mathematical Biology}
  \textbf{\bibinfo{volume}{6}}, \bibinfo{pages}{327} (\bibinfo{year}{1981}).

\bibitem[{\citenamefont{Sakaguchi and Kuramoto}(1986)}]{kuramoto}
\bibinfo{author}{\bibfnamefont{H.}~\bibnamefont{Sakaguchi}} \bibnamefont{and}
  \bibinfo{author}{\bibfnamefont{Y.}~\bibnamefont{Kuramoto}},
  \bibinfo{journal}{Pro. of Theo. Phys.} \textbf{\bibinfo{volume}{76}},
  \bibinfo{pages}{576} (\bibinfo{year}{1986}).

\bibitem[{\citenamefont{Mezard et~al.}(1987)\citenamefont{Mezard, Parisi, and
  Virasoro}}]{Mezard}
\bibinfo{author}{\bibfnamefont{M.}~\bibnamefont{Mezard}},
  \bibinfo{author}{\bibfnamefont{G.}~\bibnamefont{Parisi}}, \bibnamefont{and}
  \bibinfo{author}{\bibfnamefont{M.~A.} \bibnamefont{Virasoro}},
  \emph{\bibinfo{title}{Spin glass theory and beyond}}
  (\bibinfo{publisher}{World Scientific}, \bibinfo{year}{1987}).

\bibitem[{\citenamefont{Shiino and Fukai}(1990)}]{fukai2}
\bibinfo{author}{\bibfnamefont{M.}~\bibnamefont{Shiino}} \bibnamefont{and}
  \bibinfo{author}{\bibfnamefont{T.}~\bibnamefont{Fukai}}, \bibinfo{journal}{J.
  Phys. A} \textbf{\bibinfo{volume}{23}}, \bibinfo{pages}{L1009}
  (\bibinfo{year}{1990}).

\bibitem[{\citenamefont{Perez and Ritort}(1997)}]{Perez}
\bibinfo{author}{\bibfnamefont{C.~J.} \bibnamefont{Perez}} \bibnamefont{and}
  \bibinfo{author}{\bibfnamefont{F.}~\bibnamefont{Ritort}},
  \bibinfo{journal}{J. Phys. A} \textbf{\bibinfo{volume}{30}},
  \bibinfo{pages}{8095} (\bibinfo{year}{1997}).

\bibitem[{\citenamefont{Shiino and Fukai}(1992)}]{fukai3}
\bibinfo{author}{\bibfnamefont{M.}~\bibnamefont{Shiino}} \bibnamefont{and}
  \bibinfo{author}{\bibfnamefont{T.}~\bibnamefont{Fukai}}, \bibinfo{journal}{J.
  Phys. A} \textbf{\bibinfo{volume}{25}}, \bibinfo{pages}{L375}
  (\bibinfo{year}{1992}).

\bibitem[{\citenamefont{Aonishi et~al.}(1999)\citenamefont{Aonishi, Kurata, and
  Okada}}]{aonishi2}
\bibinfo{author}{\bibfnamefont{T.}~\bibnamefont{Aonishi}},
  \bibinfo{author}{\bibfnamefont{K.}~\bibnamefont{Kurata}}, \bibnamefont{and}
  \bibinfo{author}{\bibfnamefont{M.}~\bibnamefont{Okada}},
  \bibinfo{journal}{Phys. Rev. Let.} \textbf{\bibinfo{volume}{82}},
  \bibinfo{pages}{2800} (\bibinfo{year}{1999}).

\bibitem[{\citenamefont{Yoshioka and Shiino}(2000)}]{yoshioka}
\bibinfo{author}{\bibfnamefont{M.}~\bibnamefont{Yoshioka}} \bibnamefont{and}
  \bibinfo{author}{\bibfnamefont{M.}~\bibnamefont{Shiino}},
  \bibinfo{journal}{Phys. Rev. E} \textbf{\bibinfo{volume}{61}},
  \bibinfo{pages}{4732} (\bibinfo{year}{2000}).

\bibitem[{\citenamefont{Sompolinsky}(1987)}]{Sompolinsky}
\bibinfo{author}{\bibfnamefont{H.}~\bibnamefont{Sompolinsky}},
  \emph{\bibinfo{title}{Heidelberg Colloquium on Glassy Dynamics}}
  (\bibinfo{publisher}{Springer-Verlag}, \bibinfo{year}{1987}), pp.
  \bibinfo{pages}{485--527}.

\bibitem[{\citenamefont{Aoyagi and Kitano}(1997)}]{aoyagi2}
\bibinfo{author}{\bibfnamefont{T.}~\bibnamefont{Aoyagi}} \bibnamefont{and}
  \bibinfo{author}{\bibfnamefont{K.}~\bibnamefont{Kitano}},
  \bibinfo{journal}{Phys. Rev. E} \textbf{\bibinfo{volume}{55}},
  \bibinfo{pages}{7424} (\bibinfo{year}{1997}).

\bibitem[{\citenamefont{Okada et~al.}(1998)\citenamefont{Okada, Fukai, and
  Shiino}}]{okada}
\bibinfo{author}{\bibfnamefont{M.}~\bibnamefont{Okada}},
  \bibinfo{author}{\bibfnamefont{T.}~\bibnamefont{Fukai}}, \bibnamefont{and}
  \bibinfo{author}{\bibfnamefont{M.}~\bibnamefont{Shiino}},
  \bibinfo{journal}{Phys. Rev. E} \textbf{\bibinfo{volume}{57}},
  \bibinfo{pages}{2095} (\bibinfo{year}{1998}).

\bibitem[{\citenamefont{Daido}(1992)}]{daido}
\bibinfo{author}{\bibfnamefont{H.}~\bibnamefont{Daido}},
  \bibinfo{journal}{Phys. Rev. Let.} \textbf{\bibinfo{volume}{68}},
  \bibinfo{pages}{1073} (\bibinfo{year}{1992}).

\bibitem[{\citenamefont{Kuramoto}(1984)}]{kuramoto0}
\bibinfo{author}{\bibfnamefont{Y.}~\bibnamefont{Kuramoto}},
  \emph{\bibinfo{title}{Chemical oscillations, waves and turbulence}}
  (\bibinfo{publisher}{Springer-Verlag}, \bibinfo{year}{1984}).

\bibitem[{\citenamefont{Arenas and Vicente}(1994)}]{Vicente}
\bibinfo{author}{\bibfnamefont{A.}~\bibnamefont{Arenas}} \bibnamefont{and}
  \bibinfo{author}{\bibfnamefont{C.~J.~P.} \bibnamefont{Vicente}},
  \bibinfo{journal}{Europhys. Lett.} \textbf{\bibinfo{volume}{26}},
  \bibinfo{pages}{79} (\bibinfo{year}{1994}).

\bibitem[{\citenamefont{Bonilla et~al.}(1998)\citenamefont{Bonilla, Vicente,
  Ritort, and Soler}}]{bonilla}
\bibinfo{author}{\bibfnamefont{L.~L.} \bibnamefont{Bonilla}},
  \bibinfo{author}{\bibfnamefont{C.~J.~P.} \bibnamefont{Vicente}},
  \bibinfo{author}{\bibfnamefont{F.}~\bibnamefont{Ritort}}, \bibnamefont{and}
  \bibinfo{author}{\bibfnamefont{J.}~\bibnamefont{Soler}},
  \bibinfo{journal}{Phys. Rev. Let.} \textbf{\bibinfo{volume}{81}},
  \bibinfo{pages}{3643} (\bibinfo{year}{1998}).

\bibitem[{\citenamefont{Hoppensteadt and Izhikevich}(1999)}]{Izhikevich}
\bibinfo{author}{\bibfnamefont{F.~C.} \bibnamefont{Hoppensteadt}}
  \bibnamefont{and} \bibinfo{author}{\bibfnamefont{E.~M.}
  \bibnamefont{Izhikevich}}, \bibinfo{journal}{Phys. Rev. Let.}
  \textbf{\bibinfo{volume}{82}}, \bibinfo{pages}{2983} (\bibinfo{year}{1999}).

\bibitem[{\citenamefont{Cook}(1989)}]{cook}
\bibinfo{author}{\bibfnamefont{J.}~\bibnamefont{Cook}}, \bibinfo{journal}{J.
  Phys. A} \textbf{\bibinfo{volume}{22}}, \bibinfo{pages}{2057}
  (\bibinfo{year}{1989}).

\bibitem[{\citenamefont{Daido}(1996)}]{daido2}
\bibinfo{author}{\bibfnamefont{H.}~\bibnamefont{Daido}},
  \bibinfo{journal}{Phys. Rev. Let.} \textbf{\bibinfo{volume}{77}},
  \bibinfo{pages}{1406} (\bibinfo{year}{1996}).

\bibitem[{\citenamefont{Sherrington and Kirkpatrick}(1975)}]{sherrington}
\bibinfo{author}{\bibfnamefont{D.}~\bibnamefont{Sherrington}} \bibnamefont{and}
  \bibinfo{author}{\bibfnamefont{S.}~\bibnamefont{Kirkpatrick}},
  \bibinfo{journal}{Phys. Rev. Let.} \textbf{\bibinfo{volume}{35}},
  \bibinfo{pages}{1792} (\bibinfo{year}{1975}).

\bibitem[{\citenamefont{Aonishi}(1998)}]{aonishi}
\bibinfo{author}{\bibfnamefont{T.}~\bibnamefont{Aonishi}},
  \bibinfo{journal}{Phys. Rev. E} \textbf{\bibinfo{volume}{58}},
  \bibinfo{pages}{4865} (\bibinfo{year}{1998}).

\bibitem[{\citenamefont{Aoyagi and Kitano}(1998)}]{aoyagi3}
\bibinfo{author}{\bibfnamefont{T.}~\bibnamefont{Aoyagi}} \bibnamefont{and}
  \bibinfo{author}{\bibfnamefont{K.}~\bibnamefont{Kitano}},
  \bibinfo{journal}{Neural Computation} \textbf{\bibinfo{volume}{10}},
  \bibinfo{pages}{1527} (\bibinfo{year}{1998}).

\bibitem[{\citenamefont{Mackenzie and Young}(1982)}]{Mackenzie}
\bibinfo{author}{\bibfnamefont{N.~D.} \bibnamefont{Mackenzie}}
  \bibnamefont{and} \bibinfo{author}{\bibfnamefont{A.~P.} \bibnamefont{Young}},
  \bibinfo{journal}{Phys. Rev. Let.} \textbf{\bibinfo{volume}{49}},
  \bibinfo{pages}{301} (\bibinfo{year}{1982}).

\bibitem[{\citenamefont{Parisi}(1980)}]{parisi}
\bibinfo{author}{\bibfnamefont{G.}~\bibnamefont{Parisi}}, \bibinfo{journal}{J.
  Phys. A.} \textbf{\bibinfo{volume}{13}}, \bibinfo{pages}{L115}
  (\bibinfo{year}{1980}).

\bibitem[{\citenamefont{Mezard and Virasoro}(1985)}]{mezard2}
\bibinfo{author}{\bibfnamefont{M.}~\bibnamefont{Mezard}} \bibnamefont{and}
  \bibinfo{author}{\bibfnamefont{M.~A.} \bibnamefont{Virasoro}},
  \bibinfo{journal}{J. Physique} \textbf{\bibinfo{volume}{46}},
  \bibinfo{pages}{1293} (\bibinfo{year}{1985}).

\bibitem[{\citenamefont{Eckhorn et~al.}(1988)\citenamefont{Eckhorn, Bauer,
  Jordan, Brosch, Kruse, Munk, and Reitboeck}}]{Eckhorn1988}
\bibinfo{author}{\bibfnamefont{R.}~\bibnamefont{Eckhorn}},
  \bibinfo{author}{\bibfnamefont{R.}~\bibnamefont{Bauer}},
  \bibinfo{author}{\bibfnamefont{W.}~\bibnamefont{Jordan}},
  \bibinfo{author}{\bibfnamefont{M.}~\bibnamefont{Brosch}},
  \bibinfo{author}{\bibfnamefont{W.}~\bibnamefont{Kruse}},
  \bibinfo{author}{\bibfnamefont{M.}~\bibnamefont{Munk}}, \bibnamefont{and}
  \bibinfo{author}{\bibfnamefont{H.~J.} \bibnamefont{Reitboeck}},
  \bibinfo{journal}{Biol. Cybern.} \textbf{\bibinfo{volume}{60}},
  \bibinfo{pages}{121} (\bibinfo{year}{1988}).

\bibitem[{\citenamefont{Gray et~al.}(1989)\citenamefont{Gray, K{\"{o}}nig,
  Engel, and Singer}}]{Gray1989}
\bibinfo{author}{\bibfnamefont{C.~M.} \bibnamefont{Gray}},
  \bibinfo{author}{\bibfnamefont{P.}~\bibnamefont{K{\"{o}}nig}},
  \bibinfo{author}{\bibfnamefont{A.~K.} \bibnamefont{Engel}}, \bibnamefont{and}
  \bibinfo{author}{\bibfnamefont{W.}~\bibnamefont{Singer}},
  \bibinfo{journal}{Nature} \textbf{\bibinfo{volume}{338}},
  \bibinfo{pages}{334} (\bibinfo{year}{1989}).

\bibitem[{\citenamefont{Phillips and Singer}(1997)}]{Phillips}
\bibinfo{author}{\bibfnamefont{W.~A.} \bibnamefont{Phillips}} \bibnamefont{and}
  \bibinfo{author}{\bibfnamefont{W.}~\bibnamefont{Singer}},
  \bibinfo{journal}{Behavioral and Brain Science}
  \textbf{\bibinfo{volume}{20}}, \bibinfo{pages}{657} (\bibinfo{year}{1997}).

\bibitem[{\citenamefont{Freeman and Sharda}(1985)}]{freeman}
\bibinfo{author}{\bibfnamefont{W.~J.} \bibnamefont{Freeman}} \bibnamefont{and}
  \bibinfo{author}{\bibfnamefont{C.~A.} \bibnamefont{Sharda}},
  \bibinfo{journal}{Brain Res. Rev.} \textbf{\bibinfo{volume}{10}},
  \bibinfo{pages}{147} (\bibinfo{year}{1985}).

\end{thebibliography}

\end{document}